\documentclass[a4paper,11pt]{article}
\pdfoutput=1 

\usepackage{graphicx}  
\usepackage{dcolumn}   
\usepackage{bm,relsize}        
\usepackage{amssymb, amsmath}
\usepackage{textcomp}
\usepackage{wasysym}
\usepackage{slashed}
\usepackage{caption, subcaption}
\usepackage{multirow}
\usepackage{gensymb}

\usepackage{lipsum, color}
\usepackage[usenames,dvipsnames,svgnames]{xcolor}
\usepackage{braket}
\usepackage{jheppub} 
\usepackage{booktabs}
\usepackage{pdflscape}
\usepackage[utf8]{inputenc} 
\usepackage{booktabs}
\usepackage[left]{lineno}
\newcommand*\linenomathpatch[1]{%
  \cspreto{#1}{\linenomath}%
  \cspreto{#1*}{\linenomath}%
  \csappto{end#1}{\endlinenomath}%
  \csappto{end#1*}{\endlinenomath}%
}

\linenomathpatch{equation}
\linenomathpatch{gather}
\linenomathpatch{multline}
\linenomathpatch{align}
\linenomathpatch{alignat}
\linenomathpatch{flalign}

\def\beq{\begin{equation}}
\def\eeq{\end{equation}}
 \def\be{\begin{equation}} \def\ee{\end{equation}}
\def\bea{\begin{eqnarray}} \def\eea{\end{eqnarray}}

\newcommand{\g}{\gamma}

\newcommand{\bi}{\begin{itemize} }
\newcommand{\ei}{\end{itemize} }
\newcommand{\lp}{\left(}
\newcommand{\rp}{\right)}

\newcommand{\dd}{{\rm d}} 
\newcommand{\UN}{\vert U_N\vert}

\usepackage{caption}
\usepackage{subcaption}
\usepackage{multirow}
\usepackage[normalem]{ulem}
\allowdisplaybreaks

\usepackage[capitalise, english]{cleveref}

\AddToHook{cmd/appendix/before}{\crefalias{section}{appendix}}

\begin{document}

\title{Neutrino Transition Polarizability}
\author[a]{Matteo Borrello,}
\affiliation[a]{INFN, Sezione di Firenze Via G. Sansone 1, 50019 Sesto Fiorentino, Italy and
Department of Physics and Astronomy, University of Florence, Italy}
\author[b]{Michele Tammaro}
\affiliation[b]{Galileo Galilei Institute for Theoretical Physics, Largo Enrico Fermi 2, I-50125 Firenze, Italy}
\emailAdd{matteo.borrello@unifi.it}
\emailAdd{michele.tammaro@fi.infn.it}

\date{\today}

\preprint{}

\abstract{
We study Neutrino Transition Polarizability, where a light pseudoscalar $\phi$ couples active neutrinos to heavy neutral leptons $N$ and interacts with two photons. We consider an effective description of the active - sterile interaction, and a realization within the inverse seesaw mechanism; the coupling to photons is generated by heavy charged states. For $m_N>m_\phi$, the decay $N\to\nu\phi$ can greatly shorten the heavy neutrino lifetime, while $\phi\to\gamma\gamma$ produces a displaced photon pair. This changes the relevant search strategies, favoring production and decay inside detectors. We discuss existing constraints and estimate the sensitivity of displaced photon searches at DUNE, Belle II, LHCb and FCC-ee, highlighting the complementarity of neutrino scattering and meson and electroweak boson decays.
}

\maketitle

\section{Introduction}
\label{sec:Introduction}

Neutrinos represent a unique window into both the structure of the Standard Model (SM) and possible New Physics (NP) interactions. Their small mass hints at new mass generation mechanisms; other interactions, such as electromagnetic processes, are extremely suppressed, therefore any observation at present and planned experiments is a smoking gun signature of a new scale at play.   

Many NP theories include interactions with heavy neutral states in the neutrino sector. These Heavy Neutral Leptons (HNL) are singlets of the SM and mix with active neutrinos through the Yukawa term $y\bar L\tilde HN$, where $L$ is the left-handed lepton doublet and $H$ the Higgs doublet. Such an operator leads to a non-diagonal term in the neutrino mass matrix; upon diagonalization, the mass eigenstate of the SM neutrino receives a contribution from the HNL mass. Most notably, the seesaw mechanism \cite{Minkowski:1977sc,Mohapatra:1979ia,Foot:1988aq,Gelmini:1980re} generates the small observed neutrino mass through a very heavy Majorana mass $M_N$ of the HNL; the diagonalization leads to a light mass eigenstate of the order $m_\nu\sim y^2v^2/M_N$, where $v$ is the Higgs vev. This requires $M_N\sim10^{15}$ GeV to have $m_\nu\sim0.1$ eV, although lower masses can be achieved by asking for very small Yukawa couplings $y$~\cite{Lesgourgues:2013sjj,Lattanzi:2017ubx,Pascoli:2019wpp}. Conversely, the inverse seesaw mechanism~\cite{Mohapatra:1986bd,Gonzalez-Garcia:1988okv} relies on a small breaking of the global lepton number symmetry; the size of this breaking controls the neutrino mass through the same HNL Yukawa operator, while the mass of $N$ can naturally lie in the GeV to TeV range.

Independently of the specific neutrino mass generation mechanism, the mass matrix diagonalization leads to the mixing of HNLs with active light neutrinos; namely, $\nu\approx\nu^\prime - \sin(\alpha) N$, where $\nu$ is the light neutrino mass eigenstate, while $\nu^\prime$ and $N$ are interaction eigenstates, and $\alpha$ is the state mixing angle. Thus the HNL inherits the same weak currents as the SM neutrinos, weighted by their mass mixing, which then dictates their phenomenology. For $N$ lighter than the muon, it decays mostly to neutrinos, while above this threshold the decays are predominantly visible, hadrons and charged leptons; in both cases however, its lifetime is suppressed by the Fermi constant $G_F$, and decay lengths exceed the typical detector sizes. Searches for HNLs then rely on meson decays, looking for kinematic modifications in their spectra, or for production from protons on target, where the $N$ decays in a far detector downstream; for light masses and small mixings, cosmological observables need to be taken into account. The case for Displaced Vertices (DV) has been made only for decays at rest of heavy SM particles, such as $Z$ bosons at LEP~\cite{DELPHI:1996qcc} and $B$ mesons at Belle~\cite{Belle:2013ytx}, where the long lifetime suppression is compensated by a large statistics and low backgrounds. 

These experimental strategies are based on the idea that the neutrino sector is extended by the HNLs only. An extended Dark Sector (DS), with additional states coupling to both active and sterile neutrinos, substantially changes this picture. In this work, we explore one such generic extension of the DS, where a new scalar or pseudoscalar state $\phi$ couples the light and heavy neutrino states; we dub this interaction a Neutrino Transition. For masses of $\phi$ smaller than the HNL mass, a new decay channel opens up, $N\to\nu\phi$, with lifetimes that are orders of magnitude shorter than the mixing-induced weak decay. This Neutrino Transition alone leads to new signatures: the contribution of the $\phi$ emission affects only the neutrino neutral current (NC) scattering at leading order, effectively changing the neutral to charged current (CC) ratio\footnote{This is analogous to the effect studied by the authors in \cite{Borrello:2026bfr}, for a strongly coupled DS. In general, one can write the Neutrino Transition operator as $\Lambda^{3/2-\Delta_N}\bar LH{\cal O}_N$, where $\Lambda$ is the DS scale and ${\cal O}_N$ describes a multiparticle state in the DS. For $\Delta_N=3/2$ and ${\cal O}_N=N_R$, this reduces to the standard renormalizable HNL Yukawa interaction. Instead, considering a weakly coupled sector with ${\cal O}_N=\phi N_R$, corresponding to $\Delta_N=5/2$, one recovers the dimension five operator $\phi\bar L\widetilde H N_R/\Lambda$, which generates the NTP interaction after electroweak symmetry breaking, with $\lambda_T\sim v/\Lambda$ (see \cref{sec:Model}). Our model can therefore be regarded as a minimal, weakly coupled realization of this generalized neutrino portal. Unlike the strongly coupled scenario, where the final state can contain multiple DS particles, the additional contribution to neutral current scattering in our case involves the specific $N\phi$ final state.}.
Additionally, the $\phi$ itself can couple to the SM sector; this is typical of Axion-Like Particle (ALP) models, where $\phi$ can be coupled to two photons, e.g. through triangle loops of heavy charged particles. This setup extends the neutrino polarizability model studied in Refs.~\cite{Bansal:2022zpi,Carey:2025oxt,Gehrlein:2025tko}, and we thus call it Neutrino Transition Polarizability (NTP)\footnote{As we describe more in detail in the text, the work focuses on $m_\phi>$ MeV scale. A similar setup, focusing on $m_\phi\lesssim1$ keV and its impact on BBN, is studied in Ref.~\cite{Deppisch:2024izn}.}. Similarly, we embed this NTP into an Inverse Seesaw (ISS) model for neutrino mass generation, which naturally includes a light scalar state.

The extended DS setup of the NTP interaction leads to a distinctive phenomenology, where the $N$ can rapidly decay to a pair of visible photons through a chain of two body decays, if it is heavier than $\phi$. This can be tested across a broad range of experimental environments, exploiting the vertexing and spatial resolutions to search for two displaced photons. In this paper, we perform the first phenomenological analysis of such models, exploring the reach of both neutrino beam experiments, as well as flavor and collider experiments. In particular, we project the reach of present and future experiments for searches of DV, pointing out previously unconsidered experimental directions for HNL searches.

The paper is organized as follows. In \cref{sec:Model} we introduce the effective NTP Lagrangian, and present an explicit realization of the neutrino transition within the inverse seesaw mechanism in \cref{sec:NTPfromISS}. The main phenomenological features of NTP interactions are discussed in \cref{sec:Phenomenology}, where we examine the production and decay of the relevant degrees of freedom in \cref{sec:Production} and \cref{sec:Decay}, respectively. In \cref{sec:Results} we review the main experimental constraints on the effective NTP model and its inverse seesaw. We divide them in two categories: invisible signatures, discussed in \cref{sec:InvisibleSignals}, and visible ones, discussed in \cref{sec:VisibleSignals}. For both, we consider existing experiments as well as future probes. We present our conclusions in \cref{sec:Conclusions}.

\section{Models of Neutrino Transition Polarizability}
\label{sec:Model}
We start by building a simplified model that realizes the Neutrino Transition Polarizability (NTP). First, we consider a massive right-handed fermion, $N$, which can be either a Majorana or Dirac fermion, and a pseudoscalar $\phi$. Assuming a single generation, we have the interaction lagrangian
\beq\label{eq:Transition:Effective}
{\cal L}_{\nu N} = \lambda_T \phi \bar \nu_L N + {\rm h.c.}\,,
\eeq
which describes the light-to-heavy neutrino transition. 
In the general three-generation case, the coupling $\lambda_T$ is a complex $3\times3$ matrix, allowing for both diagonal and off-diagonal neutrino flavor transitions. 

Secondly, we allow a higher-order coupling of the pseudoscalar $\phi$ to photons, as
\beq\label{eq:Photons:Effective}
{\cal L}_{\phi\g} = \frac{g_\gamma}{4} \, \phi \, F_{\mu\nu} \widetilde{F}^{\mu\nu}\,,
\eeq
where $F_{\mu\nu}$ is the photon field strength, and 
\begin{equation}
\widetilde{F}^{\mu\nu}
\equiv
\frac{1}{2}\,\epsilon^{\mu\nu\rho\sigma}F_{\rho\sigma}
\end{equation}
is its dual. Note that the coupling $g_\gamma$ has inverse energy dimension.

The NTP is realized when both these interactions are present. Namely 
\beq\label{eq:EffectiveNTP}
{\cal L}_{\rm NTP} = {\cal L}_{\nu N} + {\cal L}_{\phi\g}\,,
\eeq
where the kinematic and mass terms are understood. This is the reference model that we consider for the phenomenological study in \cref{sec:Phenomenology,sec:Results}. The distinct feature of our analysis is the presence of $\phi$ as a light propagating degree of freedom. Indeed, we will show that the most interesting scenario is realized in the $m_\phi < m_N$ regime, allowing a new fast decay channel, $N\to\nu\phi$. Otherwise, in general we leave the two couplings, $\lambda_T$ and $g_\g$, as well as the new particle masses, $m_N$ and $m_\phi$, as free parameters. 

Beyond the simplified lagrangians, both \cref{eq:Transition:Effective,eq:Photons:Effective} can be seen as the low energy realizations of UV complete models. In this work, we focus on the model presented in Ref.~\cite{Bansal:2022zpi}, which unifies the two lagrangians into a single mechanism. On the neutrino side, the transition follows from the spontaneous breaking of the global lepton number symmetry, $U(1)_L$, in an inverse seesaw framework. The mixing induced by the Yukawa term between active and sterile neutrinos leads to the transition term, once the scalar responsible for the breaking acquires a vacuum expectation value. The two-photon coupling is generated by heavy $SU(2)_L$ charged triplets. In the minimal model, the triplets acquire their mass from the same spontaneous symmetry breaking of $U(1)_L$, responsible for the neutrino mass. In a simple non-minimal extension, the scalar that gives mass to the triplets is charged under a different $U(1)$, and the polarizability process is generated by mixing the two scalars. 
In \cref{sec:NTPfromISS}, we investigate more in depth the first part of the model, as the neutrino transition was not explicitly studied in Ref.~\cite{Bansal:2022zpi}, while we leave the second part unchanged.

As we will see in \cref{sec:Phenomenology}, the presence or absence of active--sterile mixing leads to qualitatively different phenomenological scenarios. It is however useful to remark that the neutrino transition in \cref{eq:Transition:Effective} can be generated without requiring this mixing term. A simple possibility is provided by a scotogenic-like symmetry structure~\cite{Ma:2006km,Orlofsky:2021mmy}, in which all SM fields are even under a $Z_2$ symmetry, while $N$ and the new scalar $\phi$ are odd. The usual Yukawa interaction $\bar L\widetilde H N$ is then forbidden, whereas a transition between the active neutrino and $N$ can still be generated through the $Z_2$-odd scalar sector. For an SM-singlet $\phi$, for instance, the gauge-invariant operator
\begin{equation}
    \frac{c}{\Lambda}\,\phi\,\bar L\widetilde H N
\end{equation}
generates the interaction in  \cref{eq:Transition:Effective} after electroweak symmetry breaking, with $\lambda_T=cv/(\sqrt{2}\Lambda)$~\footnote{An exact $Z_2$ symmetry would, however, also forbid the photon coupling $\phi F_{\mu\nu}\widetilde F^{\mu\nu}$. A simple way to generate it while keeping the active--sterile mixing parametrically small is to introduce a second pseudoscalar $a$, even under the $Z_2$, with
\begin{equation}
    {\cal L}\supset\frac{g_a}{4}aF_{\mu\nu}\widetilde F^{\mu\nu},
\end{equation}
and a small soft $Z_2$-breaking mixing term $\mu_{a\phi}^2 a\phi$. The mostly-$\phi$ mass eigenstate then inherits a photon coupling
\begin{equation}
    g_\gamma\simeq \theta_{a\phi}g_a,\qquad \theta_{a\phi}\simeq \frac{\mu_{a\phi}^2}{m_a^2-m_\phi^2}.
\end{equation}
Provided that neither pseudoscalar acquires a vacuum expectation value, this mixing does not generate an active--sterile mass term at tree level. The size of any radiatively induced ordinary Yukawa interaction $\bar L\widetilde H N$ then depends on the additional interactions of the UV completion and can be parametrically smaller than the NTP coupling. The limit $\mu_{a\phi}^2\to0$ restores the $Z_2$ symmetry, making the small breaking technically natural.}.

\subsection{Light to Heavy transition in Inverse Seesaw models}
\label{sec:NTPfromISS}
We start by briefly revisiting the Inverse Seesaw model described in Ref.~\cite{Bansal:2022zpi}, where the small neutrino mass follows from the small breaking of the global lepton number symmetry, $U(1)_L$. We introduce two sets of Weyl fermions, the left- and right-handed $N_L^i$ and $N_R^i$, each carrying lepton number $L=+1$ and generation index $i=1,2,3$, and the scalar $S$, with $L=-2$.

The interaction lagrangian reads
\begin{equation}\label{eq:ISS}
    -{\cal L}=y_\nu\bar L \tilde H N_R+\bar N_L M_N N_R+\frac{\lambda_R}{2}\bar N_R^c SN_R+\frac{\lambda_L}{2}\bar N_L^c SN_L+h.c.\,,
\end{equation}
where $L_i$ are the SM lepton doublets, and $H$ is the SM Higgs doublet. The Yukawa coupling $y_\nu$ and neutrino Dirac mass $M_N$ are general $3\times3$ complex matrices, while $\lambda_{L,R}$ are symmetric $3\times3$ complex matrices.  For simplicity, we focus again on the one-generation case, and refer to the discussion in Ref.~\cite{Bansal:2022zpi} for a generalization to the three-family case. In this case, by field redefinitions $y_\nu$, $M_N$ and $\lambda_L$ can be made real, while $\lambda_R$ remains complex.

The lepton number is spontaneously broken once $S$ obtains a vev
\begin{equation}
    S=\frac{f_\phi+\sigma+i\phi}{\sqrt{2}}\,,
\end{equation}
where $f_\phi$ is the vev, $\sigma$ is the radial mode and $\phi$ is the pNGB of the $U(1)_L$ symmetry. The radial mode acquires a mass $m_\sigma\sim {\cal O}(f_\phi)$\footnote{The statement that $m_\sigma$ is set by $f_\phi$ is only parametric. For example, for 
$$V(S) = \lambda_S \left( |S|^2 - \frac{f_\phi^2}{2} \right)^2,$$ 
one obtains 
$$m_\sigma^2 = 2\lambda_S f_\phi^2.$$ 
Thus $m_\sigma \sim f_\phi$ only if $\lambda_S = \mathcal{O}(1)$.}, while the pNGB mass $m_\phi$ represents a small explicit breaking of the $U(1)_L$, ${\cal L} \supset-m_\phi^2\phi^2/2$, and is taken as a free parameter.

After electroweak symmetry breaking, the neutrino sector acquires a non-diagonal mass matrix. In the $\left\{\nu_L^c,N_R,N_L^c\right\}$ basis, it reads
\begin{equation}
    \mathcal{M} =
\begin{pmatrix}
0 & M_D & 0 \\
M_D & \mu_R & M_N \\
0 & M_N & \mu_L
\end{pmatrix}\,,
\end{equation}
where we define the Dirac mass $M_D=y_\nu v/\sqrt2$ and the parameters $\mu_{L,R}=\lambda_{L,R}f_\phi/\sqrt2$; we also assume the hierarchy $\mu_{R,L}\ll M_D\ll M_N$. We find the lightest neutrino mass eigenstate to be
\beq
\nu \simeq c_\alpha\nu_L - s_\alpha N_L\,,
\eeq
with higher order corrections proportional to $\mu_{L,R}/M_{D,N}$. Here we have used the abbreviations $\sin\alpha\equiv s_\alpha$, $\cos\alpha\equiv c_\alpha$, and we have defined the mixing angle $\alpha$ as
\beq
 \tan\alpha \equiv U_N \simeq \frac{M_D}{M_N}\,.
\eeq
The corresponding mass eigenvalue reads
\begin{equation}\label{eq:LightNeutrinoMass}
    m_\nu \simeq \mu_L \UN^2\,,
\end{equation}
while the two heavy mass eigenstates, $N_{1,2}$, have nearly degenerate masses of order $m_N\sim\sqrt{M_D^2 + M_N^2}$.
Thus the smallness of the neutrino mass is given by either a small mixing angle, a small vev $f_\phi$, a small $\lambda_L$, or a combination of the three; it vanishes in the limit $f_\phi\to0$, that is in the absence of a spontaneous breaking of the lepton number symmetry. The coupling $\lambda_L$, which governs the Yukawa term between $N_L$ and the scalar $S$, is not constrained by any symmetry argument and is naturally of ${\cal O}(1)$. In the remaining of the paper, we fix $|\lambda_L|=1$.

We can now write down the couplings of the radial and longitudinal excitations, $\sigma$ and $\phi$, of the scalar field $S$. These are obtained by replacing $\mu_{L,R}\to\mu_{L,R}\lp 1 + \frac{\sigma + i\phi}{f_\phi} \rp$, and then rotating to the mass basis. We explicitly rewrite the interaction eigenstates as a combination of the mass ones; up to small corrections of ${\cal O}(\mu_{L,R})$, we have  
\begin{align}
    \nu_L&\approx \nu +\frac{\UN}{\sqrt 2}(N_2+N_1)\,,\\
    N_L&\approx\frac{N_1+N_2}{\sqrt 2}-\UN\nu\,,\\
    N_R&\approx-\frac{N_1-N_2}{\sqrt 2}\,.
\end{align}
The couplings of the two modes will be the same, while only the Lorentz structure will differ, scalar for $\sigma$ and pseudoscalar for $\phi$.
The interaction of $\phi$ to light neutrinos is
\beq
{\cal L}_{\nu\nu} = -\frac{i}{2}\frac{m_\nu}{f_\phi}\phi\lp\bar\nu P_L \nu\rp = -\frac{i}{2\sqrt2} \UN^2 \phi\lp\bar\nu P_L \nu\rp\,,
\eeq
thus the coupling reads
\beq\label{eq:couplingLight}
c_\nu =-\frac{i}{\sqrt 2}\UN^2\,.
\eeq
As already noted in Ref.~\cite{Bansal:2022zpi}, this coupling can be ${\cal O}(1)$ if $f_\phi\sim m_\nu$, although phenomenological constraints exclude such large values.

Secondly, the light-to-heavy transition lagrangian is 
\begin{equation}
    -{\cal L}_{\rm \nu N} = -\frac{i}{2}\UN\phi\lp\bar N_1^c+\bar N_2^c\rp\nu+h.c.\,,
\end{equation}
which can be matched directly onto \cref{eq:Transition:Effective}, leading to 
\beq\label{eq:lambdaTtoMixing}
\lambda_T = \frac{\UN}{\sqrt 2}\,,
\eeq
where we identified for simplicity the field $N$ in \cref{eq:Transition:Effective} with $(N_1+N_2)/\sqrt 2$. 
That is, the transition is at leading order proportional to the mixing angle $U_N$, as expected from the fact that there are no other mechanisms to couple the two states beyond the Yukawa term. Note that in the 3-generation case, both $c_\nu$ and $U_N$ are generic 3$\times$3 complex matrices, and can inherit flavor diagonal and off-diagonal elements. The flavor structure plays a significant role when weak processes dominate, as in the benchmark in \cref{sec:otherbenchmarks}, whereas the NTP-dominated scenario discussed in \cref{sec:Results} is largely insensitive to it. Throughout our analysis, we assume for simplicity that the HNL mixes exclusively with the muon neutrino. 

Finally, we remind that the active - sterile mixing leads to a modification of the weak currents. Moving from the flavor basis to the mass one, one obtains for charged currents
\beq\label{eq:WeakMixing}
{\cal L}_{\rm weak} \supset \frac{g_2}{\sqrt2}W_\mu^- \bar\ell \gamma^\mu P_L \lp \nu + U_N N \rp\,,
\eeq
and similarly for neutral currents. Here $g_2$ is the $SU(2)_L$ gauge coupling, and we are assuming again for simplicity to have only one lepton generation. Thus, in the inverse seesaw model presented here, as well as any model that features the Yukawa term as in \cref{eq:ISS}, a light-to-heavy neutrino transition is always present through weak currents.

The relevant parameters of this model are the lepton number breaking scale, $f_\phi$, and the Dirac and Majorana masses of the sterile states, $M_D$ and $M_N$, whose ratio gives the mixing angle, $U_N\simeq M_D/M_N$. 
Requiring that \cref{eq:LightNeutrinoMass} reproduces the typical value of neutrino masses, $m_\nu = 0.1$ eV, we have
\beq
f_\phi\approx \sqrt 2 m_\nu \frac{m_N^2}{M_D^2} \sim 100~{\rm keV}\times\lp \frac{m_\nu}{0.1~{\rm eV}} \rp\times\lp \frac{m_N}{{\rm GeV}} \rp^2 \times \lp \frac{{\rm MeV}}{M_D} \rp^2\,,
\eeq
The neutrino mass can then be recovered with a small sub-MeV breaking scale and a mixing angle of $\UN^2\simeq10^{-6}$, if we consider an HNL with mass in the GeV range. This translates directly into a requirement on the coupling of $\phi$ to neutrinos, as $c_\nu\sim\UN^2$, as shown in \cref{eq:couplingLight}.

However, we can obtain a stricter relation by requiring a good approximate $U(1)_L$ symmetry, that is by asking that the explicit symmetry breaking term, $m_\phi$, is smaller than the spontaneous one, $f_\phi$. We have
\begin{equation}
  f_\phi\sim \frac{m_\nu}{\UN^2}\gtrsim m_\phi\implies \UN^2\lesssim\frac{m_\nu}{m_\phi}
\end{equation}
which leads to 
\begin{equation}\label{eq:numassregion}
\UN^2
\lesssim
10^{-7}
\left(\frac{m_\nu}{0.1~\mathrm{eV}}\right)
\left(\frac{1~\mathrm{MeV}}{m_\phi}\right)\,.
\end{equation}
This relation is more relevant for the phenomenology of the NTP studied in this manuscript, as the parameter space of pseudoscalars lighter than a few MeV is largely excluded by cosmological and astrophysical observables, while heavier states are accessible at present and future terrestrial experiments~\cite{Bansal:2022zpi,Carey:2025oxt}.

\section{Phenomenology of NTP}
\label{sec:Phenomenology}
In this Section, we compute the production and decay rates of the HNL state, $N$, for both the effective NTP lagrangian, \cref{eq:EffectiveNTP}, and for the ISS model in \cref{sec:NTPfromISS}, as well as those for the pseudoscalar $\phi$. The main difference between the two models is the modification of weak currents, \cref{eq:WeakMixing}, that is present only in the inverse seesaw model. We derive expressions for both channels\footnote{A thorough review of HNL phenomenology through the active-sterile neutrino mixing in weak currents is given in Ref.~\cite{Bondarenko:2018ptm}, which we refer to for the rates shown in this section.}. 

\subsection{Production}
\label{sec:Production}

\begin{figure}[t!]
    \centering
    \includegraphics[width=0.4\linewidth]{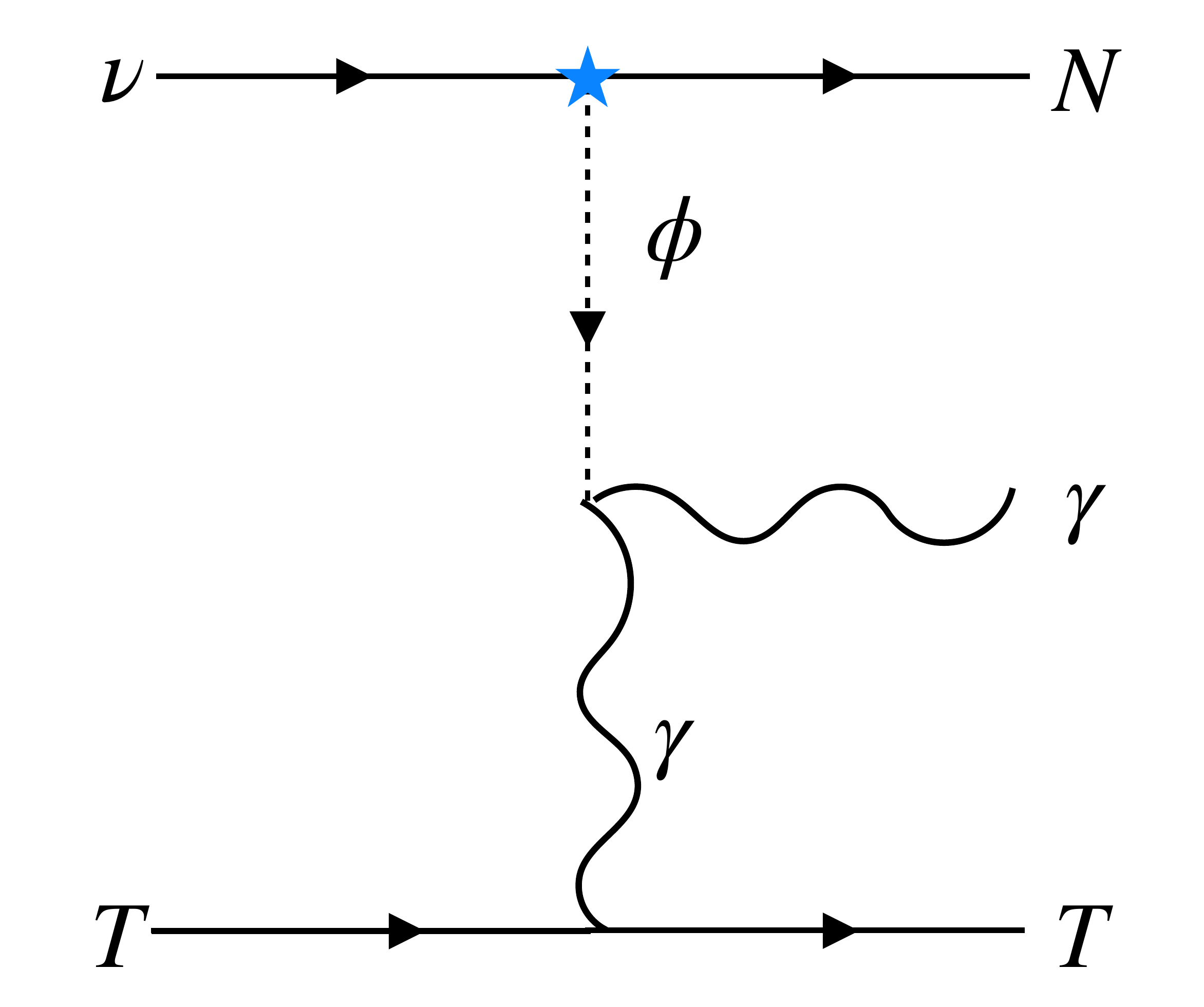}\hspace{2cm}
    \includegraphics[width=0.4\linewidth]{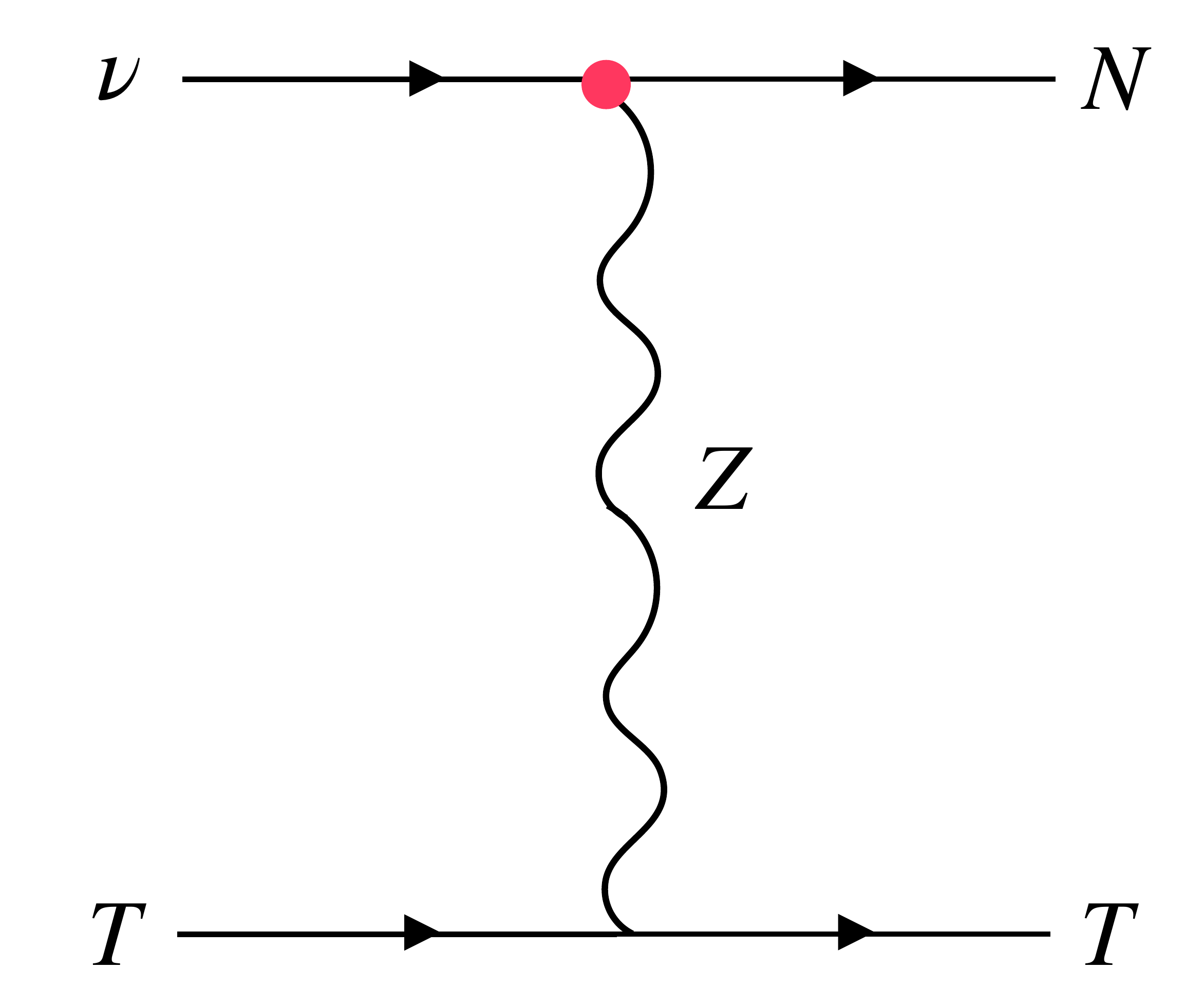}
    \caption{Production of a heavy neutral lepton $N$ through neutrino scattering on a target $T$. Here and in the following, the red dot denotes a weak-mixing vertex, while the blue star denotes the NTP insertion. {\bf Left:} NTP-induced scattering accompanied by photon emission in the effective model described by \cref{eq:EffectiveNTP}. {\bf Right:} Standard weak scattering NC event mediated by $Z$-boson exchange and induced by the active--sterile neutrino mixing of \cref{eq:WeakMixing}. }
    \label{fig:ntp}
\end{figure}

\paragraph{Electromagnetic neutrino scattering:} The NTP lagrangian, \cref{eq:EffectiveNTP}, allows for $N$ to couple directly with the active neutrinos $\nu$ through $\phi$, which in turn can couple directly to photons. This is shown in the left diagram of \cref{fig:ntp}.
The process of interest is $\nu + T\to N + \g + T$\footnote{The $2\to2$ process $\nu + T\to N + T$ can also be generated, by attaching both photons to the target. The one-loop diagram is expected to scale as $e^4Z^4g_\g^2\lambda_T^2/(4\pi)^2$, while the tree-level $2\to3$ scatterings as $e^2Z^2g_\g^2\lambda_T^2$. Their ratio then is always much smaller than one, and we neglect this contribution in the rest of the discussion.}, where $T$ is a charged target, either an electron or a nucleus. In the latter case, the dynamics of the scattering depends on the momentum exchanged by the virtual photon, $q_\g^2$. For small enough momenta, $q_\g\lesssim1/R_A$, where $R_A$ is the typical radius of the nuclear target, the neutrino can only resolve the entire nucleus; thus the scattering is coherently enhanced by the total nuclear charge, $Z$. For larger values, $q_\g^2\gtrsim\Lambda_{\rm QCD}^2$, the neutrino scatters on the single nucleons, ejecting them from the nucleus; for even larger values, the neutrino undergoes Deep Inelastic Scattering (DIS) with partons in the nucleons. 
In the following, we focus on coherent scattering of neutrinos on nuclei, as this is the most relevant process in the experimental setups studied in \cref{sec:Results}.

The differential cross section of the process reads
\beq\label{eq:NTP:nucleus:xs}
\dd\sigma =
\frac{1}{4E_\nu m_A}\vert\overline{\cal M}\vert^2
\dd\Phi_3\,,
\end{equation}
where we have taken the initial nucleus to be at rest and neglected the active-neutrino mass; here $d\Phi_3$ is the three-body final state phase space, and $\overline{\cal M}$ denotes the amplitude summed over final-state polarizations and nuclear spin.
The squared amplitude reads 
\beq\label{eq:NTP:nucleus:Msq}
|\overline{\cal M}|^2 = |F(Q_\g^2)|^2 |\lambda_T g_\g Z e|^2 \frac{\left(Q_\phi^2+m_N^2\right)E_\g^2} {\left(Q_\phi^2+m_\phi^2\right)^2} \frac{2m_T+T_T}{8T_T} \sin^2\theta_{\g T}\,,
\eeq
where $Q_\phi^2\equiv-q_\phi^2$, with $q_\phi=p_\nu-k_N$, while $Q_\g^2\equiv-q_\g^2=2m_TT_T$ is the virtuality of the photon exchanged with the target. Here, $T_T$ is the nuclear recoil energy and $\theta_{\g T}$ denotes the angle between the emitted photon and the nuclear recoil momentum. The angular dependence can equivalently be expressed in terms of the angle $\theta_{\g N}$ between the photon and the heavy state. Using momentum conservation, $\mathbf{k}_T=\mathbf{p}_\nu-\mathbf{k}_N-\mathbf{k}_\g$, one finds
\beq 
\frac{2m_T+T_T}{8T_T}\sin^2\theta_{\g T}=\frac{T_T(2m_T+T_T)-\left(E_\nu\cos\theta_\g-|\mathbf{k}_N|\cos\theta_{\g N}-E_\g\right)^2}{8T_T^2}\,, 
\eeq
where $\theta_\g$ is the angle between the emitted photon and the incoming neutrino direction, $|\mathbf{k}_N|=\sqrt{E_N^2-m_N^2}$ is the three-momentum of the heavy state, and $E_N=E_\nu-E_\g-T_T$ follows from energy conservation in the target rest frame.
Relative to the light-to-light process of Ref.~\cite{Carey:2025oxt}, the fermionic current amounts to the replacement $Q_\phi^2\to Q_\phi^2+m_N^2$. The function $F(Q_\g^2)$ is the electromagnetic nuclear form factor and accounts for the loss of coherence at finite momentum transfer. We consider the Helm form factor~\cite{Duda:2006uk} (see~\cite{Carey:2025oxt} for a discussion on other possible form factors).
 
To derive the energy and angular distributions of $N$ and $\g$, we implement the NTP model in {\tt MadGraph v3.7.0}~\cite{Alwall:2014hca}, to simulate the coherent process $\nu {\rm T}\to N\g {\rm T}$, where $T$ is a nucleus target. We focus on scattering on Argon ($Z_T=18$, $A_T=40$), as the relevant neutrino beam experiment is the DUNE Near Detector (see \cref{sec:Results}).
We scan over the incoming neutrino energy in the range $E_\nu\in[1,10]~{\rm GeV}$, in steps of $0.5~{\rm GeV}$, and consider HNL masses $m_N=(1,3,5,8)\times10^i~{\rm GeV}$, with $ i=-2,-1,0$. For each $(E_\nu,m_N)$ point, we generate $10^5$ events, imposing a low energy cut $p_{T\g}>1~{\rm MeV}$ to improve the numerical stability of the phase-space integration. 
The events are generated for the reference value $m_\phi=50~{\rm MeV}$; results for different mediator masses are then obtained by reweighting each event according to the corresponding $\phi$ propagator in \cref{eq:NTP:nucleus:Msq}: to an event generated at the reference mass $m_{\phi,0}=50~{\rm MeV}$, we assign a weight
\beq
\left(
\frac{Q_\phi^2+m_{\phi,0}^2} {Q_\phi^2+m_\phi^2}\right)^2\,.
\eeq
The kinematical properties of the signal are discussed in \cref{sec:DUNE} for the $m_\phi=50$ MeV benchmark.

\begin{figure}[t!]
    \centering
    \includegraphics[width=0.7\linewidth]{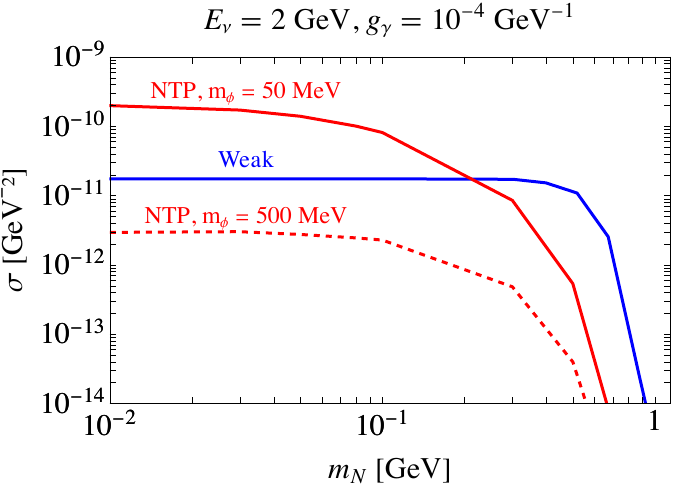}
  \caption{Coherent production cross sections of a heavy neutral lepton $N$ in neutrino scattering on Argon, as a function of $m_N$, for an incoming neutrino energy $E_\nu=2$ GeV. The red curves show NTP-induced production for $m_\phi=50$ MeV (solid) and $500$ MeV (dashed), with $g_\gamma=10^{-4}~{\rm GeV}^{-1}$. The blue curve corresponds to weak up-scattering through active--sterile mixing. In both cases we are assuming $|U_N|^2=1$.}

    \label{fig:nuxsplot}
\end{figure}

\paragraph{Weak neutrino scattering:} In the ISS model, heavy states can be produced also by coherent up-scattering through the process $\nu +T\to N+T$, mediated by a $Z$ boson, where the $N$ mixes with $\nu$ on the external outgoing leg; this is shown in the right diagram of \cref{fig:ntp} \footnote{Note that at charged lepton beam experiments, a similar process can happen via charged current scattering only in the inelastic regime, as $\ell +T\to N+T^\prime$, where $T^\prime$ is an excited nuclear state, or a lower mass nucleus plus an ejected nucleon.}.
The differential cross-section for coherent up-scattering production of heavy states mediated by $Z$ boson exchange is
\beq\label{eq:weakxs}
\frac{\dd \sigma_{ w}}{\dd Q^2} =\frac{\vert \overline{{\cal M}_w}\vert^2}{64\pi m_T^2 E_\nu^2}\,.
\eeq
Here $Q^2=-q^2>0$ denotes the squared momentum transfer. For a spin-zero
nuclear target, the squared matrix element is
\beq
\vert \overline{{\cal M}_w}\vert^2=\frac{G_F^2 |U_N|^2 Q_w^2 |F_w(Q^2)|^2}{2}\left[ \left(4 m_T E_\nu - m_N^2 - Q^2\right)^2 - \left(4 m_T^2 + Q^2\right) \left(m_N^2 + Q^2\right) \right]\,,
\eeq
where $Q_w = Z_T(1 - 4\sin^2\theta_w) - N_T$ is the effective $Z$ boson coupling to $Z_T$ protons and $N_T$ neutrons, while $\theta_w$ is the weak mixing angle and we consider an Argon target~\footnote{The effective lagrangian is \beq
{\cal L}=\frac{G_F}{\sqrt 2}U_N Q_w \lp\bar N \gamma_\mu P_L \nu\rp\lp T^\dagger i \overleftrightarrow{\partial^\mu} T\rp\,.
\eeq}, with $m_T\simeq40$ GeV. The function $F_w(Q^2)$
is the corresponding weak nuclear form factor, which we take from Ref.~\cite{VanDessel:2020epd}.

For $Q^2\ll m_Z^2$, as expected in the coherent regime, the cross section is proportional to the Fermi coupling constant $G_F^2\propto m_Z^{-4}$. In the limit $m_N\ll E_\nu\ll m_T$, the differential cross-section takes the simple form
\beq
\frac{d\sigma}{dQ^2}\simeq\frac{G_F^2 |U_N|^2 Q_w^2}{8\pi}|F_w(Q^2)|^2\,.
\eeq
The total coherent rate is therefore controlled by the range of momentum transfers over which the weak nuclear form factor remains sizeable.

In \cref{fig:nuxsplot} we compare the electromagnetic and weak $N$ production cross sections, for a fixed incoming neutrino energy, $E_\nu=2$ GeV, and two different $\phi$ masses, $m_\phi=50,500$ MeV; for the former process, we assume $g_\g=10^{-4}~{\rm GeV}^{-1}$. In the lighter $m_\phi$ benchmark, the weak production rate is smaller by an order of magnitude than the NTP one at low $m_N$; however, the different mass scaling makes the NTP cross section decrease faster, so that weak production eventually becomes relevant at $m_N\sim1$ GeV. For the $m_\phi=500$ MeV benchmark, the weak production is always dominant by at least an order of magnitude over the entire $m_N$ range.

\begin{figure}[t!]
    \centering
    \includegraphics[width=0.49\linewidth]{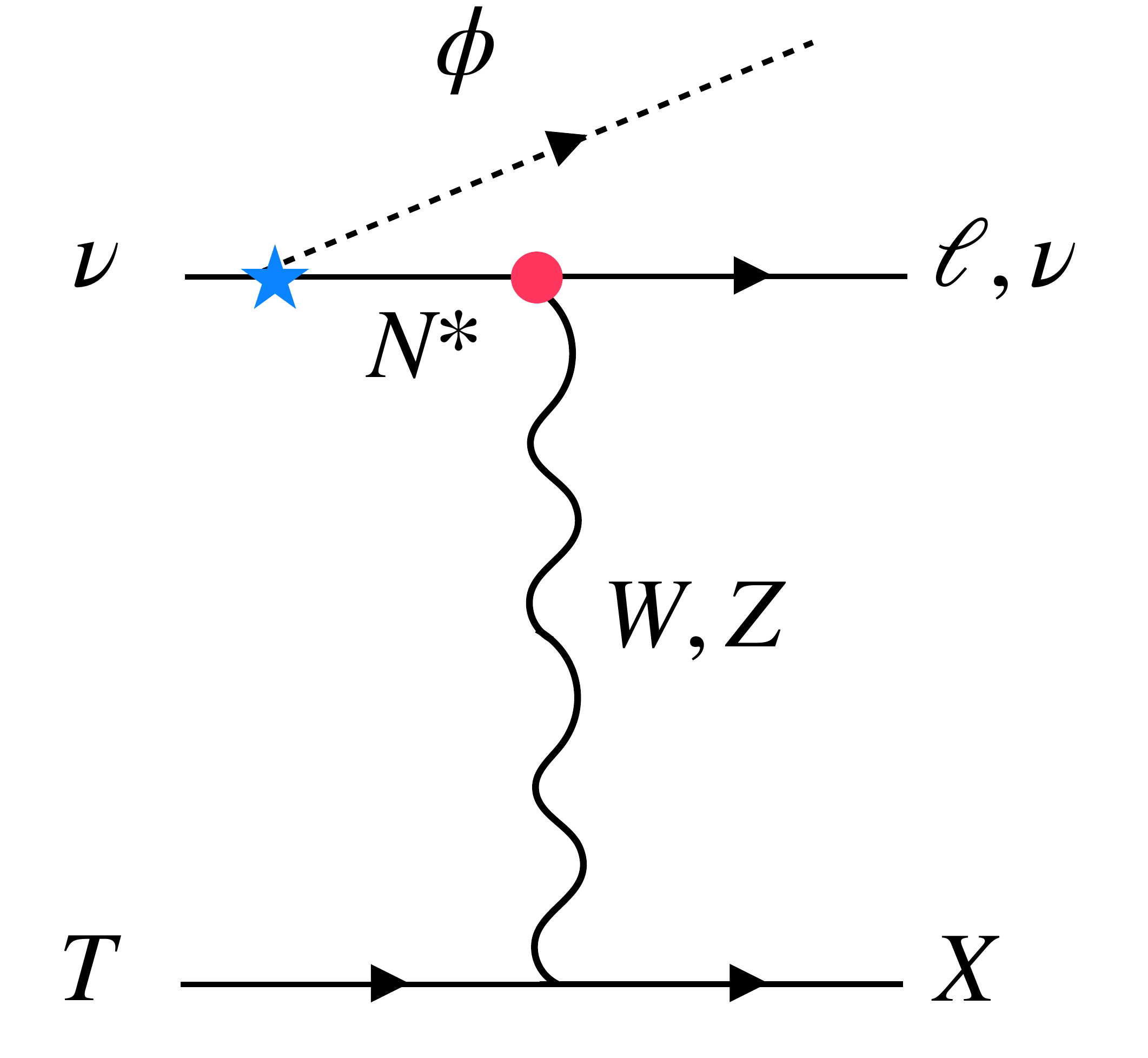}
        \includegraphics[width=0.49\linewidth]{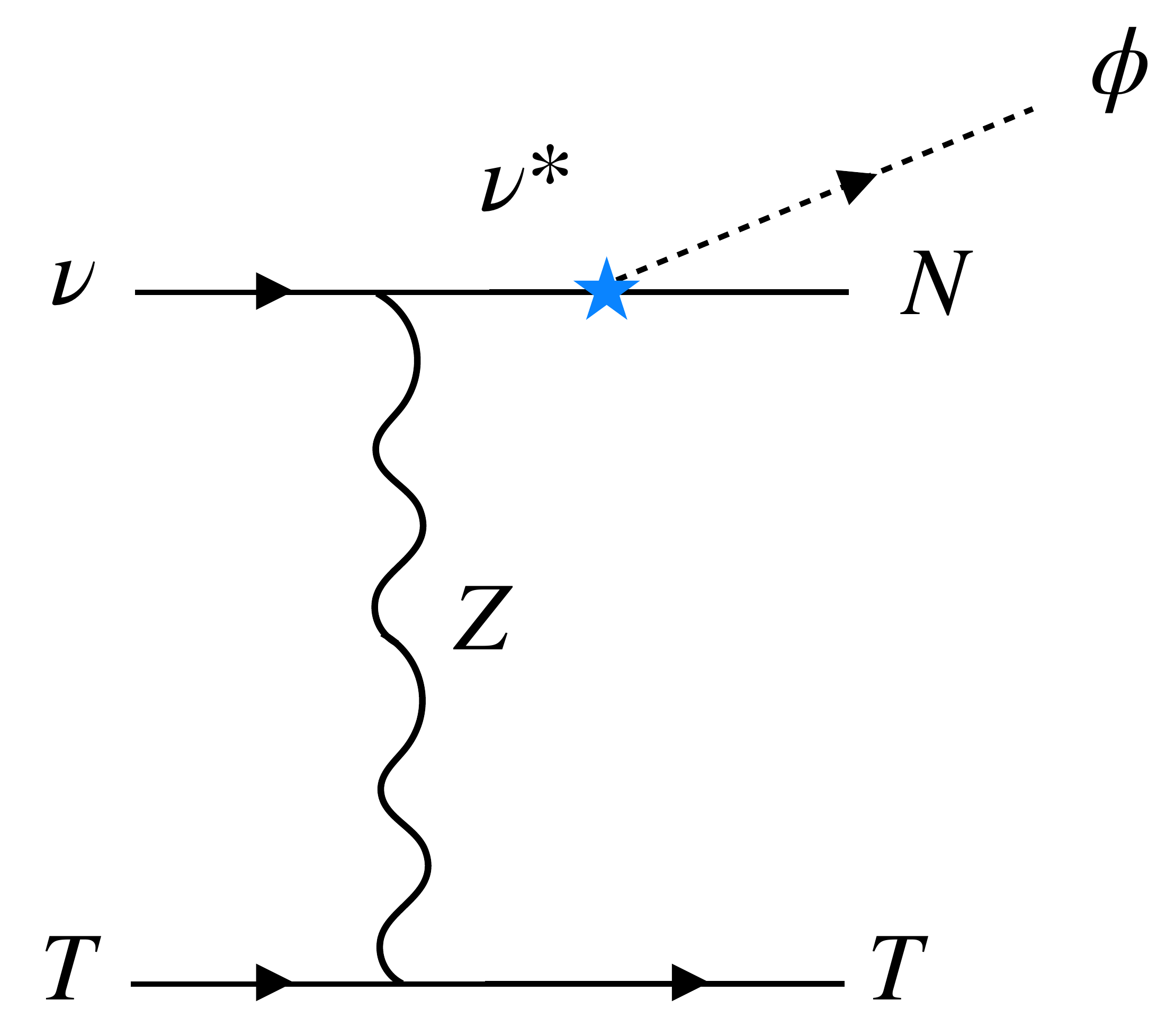}
\caption{Radiative emission of the pseudoscalar $\phi$ in neutrino scatterings. \textbf{Left:} Charged- or neutral-current production, $\nu T\to \ell(\nu)\phi X$, mediated by an off-shell HNL $N^*$ and requiring active--sterile mixing. \textbf{Right:} Neutral-current production, $\nu T\to N\phi T$, mediated by an off-shell active neutrino $\nu^*$ and involving only the SM neutral-current interaction and the NTP vertex.}
    \label{fig:nccc:enhancement}
\end{figure}

\paragraph{Radiative $\phi$ emission:} The pseudoscalar $\phi$ can be radiated from the neutrino and $N$ legs in weak scattering processes, shown in \cref{fig:nccc:enhancement}. However, note that the process $\nu T\to N\phi T$ (right diagram) in generic NTP models will affect only the NC process, as it requires an off-shell light neutrino in the outgoing leg, while it cannot enter the CC. This effect leads to enhanced inclusive NC/CC ratio, as discussed in \cref{sec:InvisibleSignals}.

A naive estimate of the rate can be obtained by treating the emission of $\phi$ as a radiative splitting of the neutrino line. For NC scattering one expects
\beq\label{eq:neutrinothreebody}
\frac{\sigma_{\rm NC+\phi}}{\sigma_{\rm NC}^{\rm SM}}\sim \frac{|\lambda_T|^2}{16\pi^2}L\,,
\eeq
up to ${\cal O}(1)$ phase-space factors. The logarithmic IR enhancement $L$ is 
\beq\label{eq:splitting}
\log\left[\frac{Q^2}{(m_N+m_\phi)^2}\right]\,,
\eeq
and is associated with the nearly collinear splitting $\nu^*\to N\phi$~\footnote{Schematically, the logarithmic enhancement is computed as
\begin{equation*}
L\sim \int dz\,P_{\nu\to N\phi}(z) \log\left[ \frac{Q_{\rm hard}^2} {(1-z)m_N^2+z m_\phi^2} \right]\,,
\end{equation*}
where $z$ denotes the momentum fraction carried by $N$. Neglecting mass effects in the splitting kernel $P_{\nu\to N\phi}$, one has $P_{\nu\to N\phi}(z)\propto (1-z)$. Note that, unlike bremsstrahlung of gauge bosons, scalars do not exhibit a soft divergence. The scale $Q_{\rm hard}$ sets the upper virtuality available, and it is parametrically of the order of the weak momentum transfer, $Q_{\rm hard}^2\sim Q^2=-q_Z^2$. For an order-of-magnitude estimate, the logarithmic enhancement may therefore be represented as $L\sim\log[Q_{\rm hard}^2/(m_N+m_\phi)^2]$, up to ${\cal O}(1)$ factors.}.

\begin{figure}[t!]
    \centering
    \includegraphics[width=0.49\linewidth]{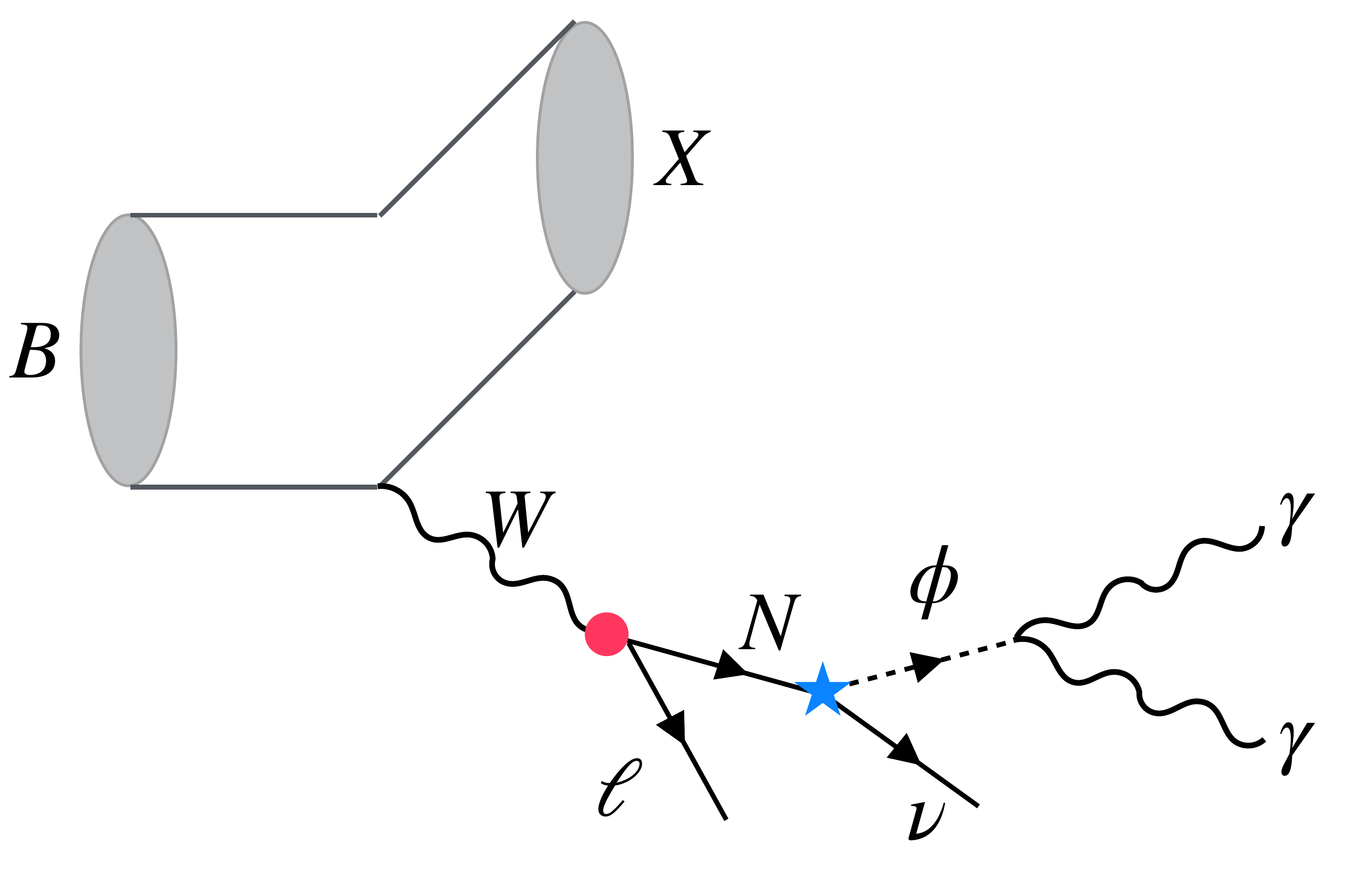}
    \includegraphics[width=0.49\linewidth]{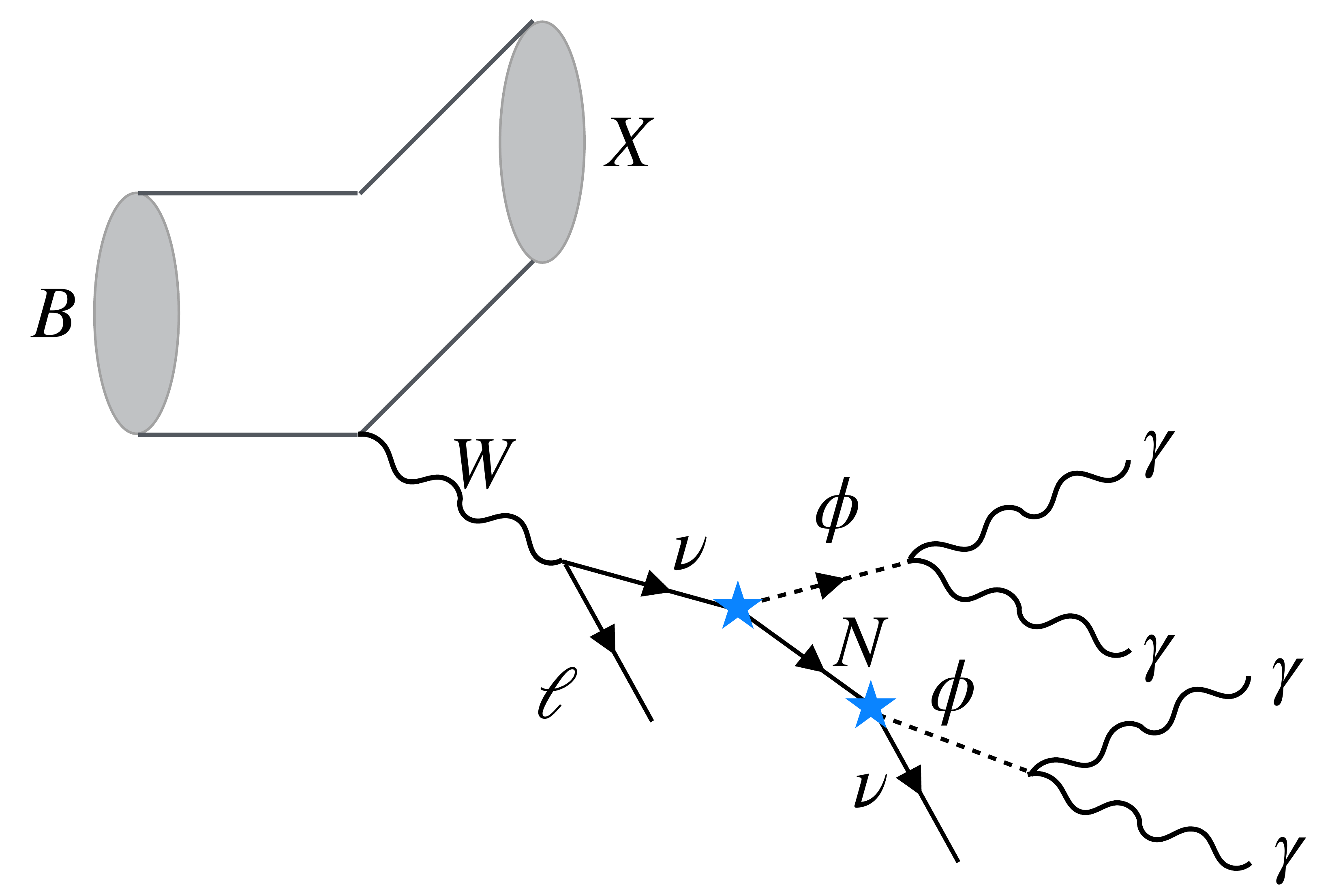}
    \includegraphics[width=0.45\linewidth]{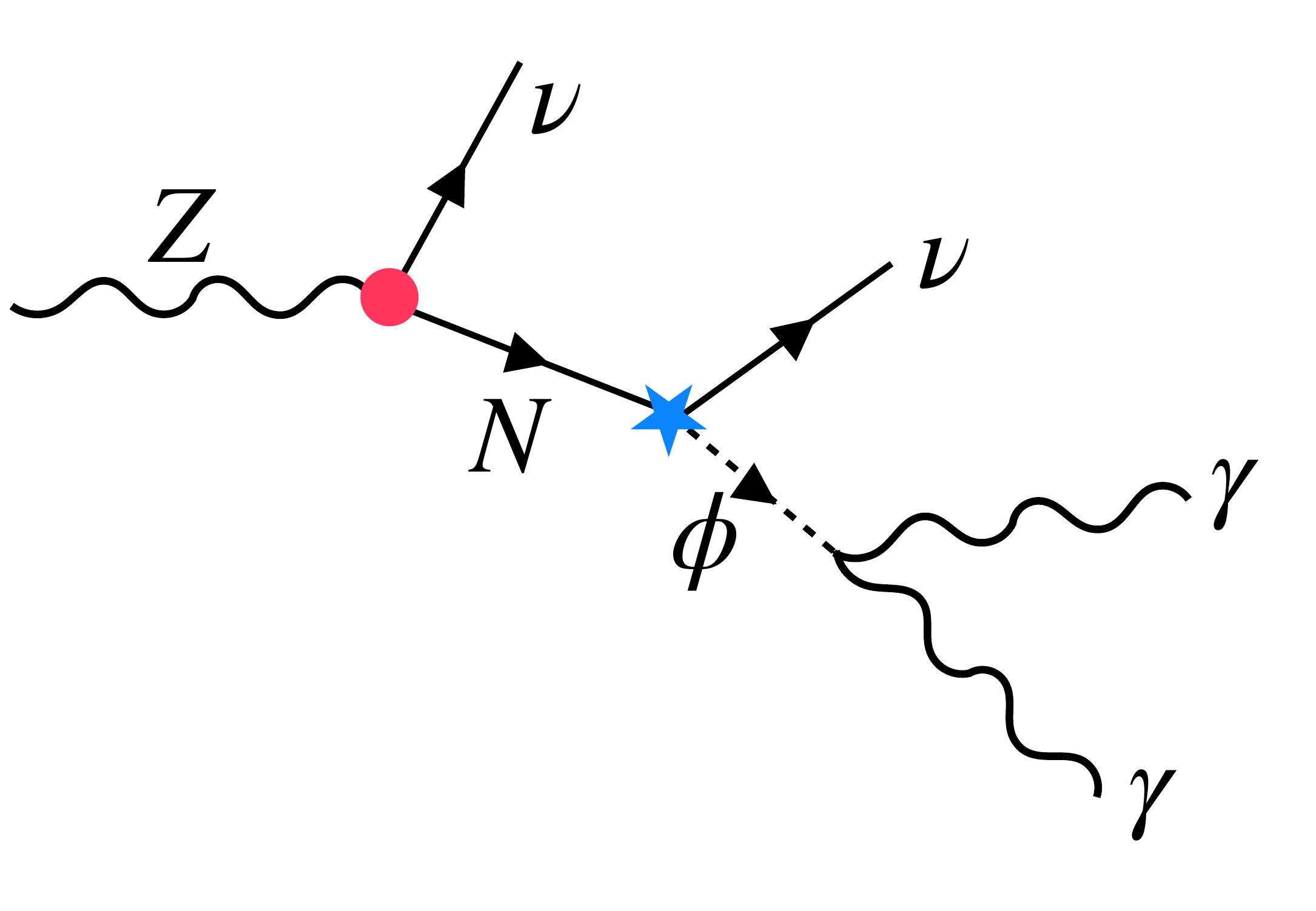}
    \includegraphics[width=0.45\linewidth]{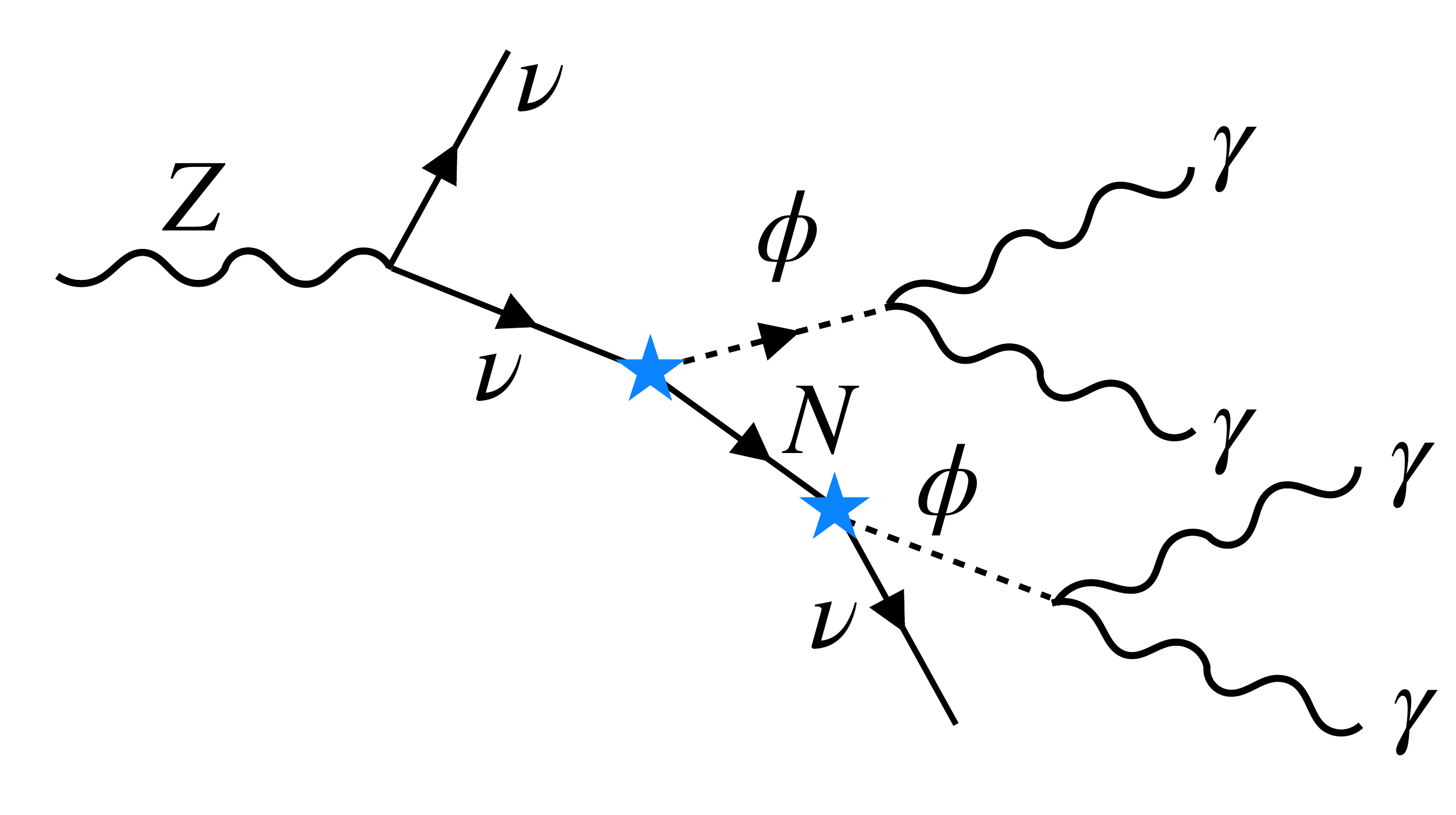}
    \caption{Topologies of meson and boson decays to $N$ and $\phi$. \textbf{Top left:} HNL production through active--sterile mixing in a semileptonic $B$-meson decay, followed by the NTP decay $N\to\nu\phi$ and $\phi\to\gamma\gamma$. \textbf{Top right:} HNL production through splitting of the off-shell neutrino line, $\nu\to N\phi$, followed by the cascade $N\to\nu\phi$, yielding two scalar mediators and hence up to four photons in the final state. \textbf{Bottom:} Same for the case of $Z$ boson decays. 
    }
    \label{fig:bosproduction}
\end{figure}

\paragraph{Heavy SM particle decays:} A different production channel for $N$ states is the decay of SM particles, as gauge bosons and mesons, shown in \cref{fig:bosproduction}. The left top and bottom diagrams correspond to the decay through light-heavy neutrino mixing, 
\beq\label{eq:MixingDecays}
B\to X\ell N\,, \qquad Z\to \nu N\,, 
\eeq
and are thus proportional to $\UN^2$. We refer to \cite{Bondarenko:2018ptm} for these production rates. If kinematically allowed, $N$ can subsequently decay to $\phi$; in the opposite regime, direct production of $\phi$ requires an off-shell HNL, $N^*\to\nu\phi$, which we estimate to be negligible, and do not consider them further. 

As in the case of radiative $\phi$ emission, the NTP interaction opens a genuinely new production topology, where $N$ and $\phi$ are attached to an off-shell neutrino leg; an example is shown in the right top and bottom diagrams of \cref{fig:bosproduction}.  
The resulting processes, 
\beq\label{eq:NoMixingDecays}
B\to X\ell N\phi\,, \qquad Z\to \nu N \phi\,, 
\eeq 
require one insertion of $\lambda_T$, and an additional body in the final state.
A simple parametric estimate can be obtained away from kinematic thresholds, by considering the process as an ordinary SM decay into an off-shell active neutrino, followed by the splitting $\nu^*\to N\phi$. We have
\beq \label{eq:decaywidthmodification}
\frac{\Gamma_{\rm NTP}}{\Gamma_{\rm SM}}\sim\frac{|\lambda_T|^2}{16\pi^2}L\sim{\cal O}(10^{-2})|\lambda_T|^2\,, 
\eeq
where $L$ parametrizes the logarithmic enhancement associated with the splitting, as defined in \cref{eq:splitting}. For comparison, the mixing suppressed decays in \cref{eq:MixingDecays} scale as
\beq 
\frac{\Gamma_{\rm mix}}{\Gamma_{\rm SM}}\sim |U_N|^2\,\rho(m_N)\,,
\eeq
where $\rho(m_N)$ denotes the corresponding phase space factor. In the ISS realization considered here, $\lambda_T=|U_N|/\sqrt2$, so that both production mechanisms inherit the same overall $|U_N|^2$ suppression. Their ratio is therefore parametrically
\beq\label{eq:mixingvsntp} 
\frac{\Gamma_{\rm mix}}{\Gamma_{\rm NTP}}\sim{\cal O}(10^3)\rho(m_N)\,. 
\eeq 
Thus, genuinely NTP-induced production is typically subleading to standard mixing-induced production in the ISS realization.

\subsection{Decay}
\label{sec:Decay}
The kinematics of the decay chain depends on the mass hierarchy between the heavy neutrino $N$ and the pseudoscalar $\phi$. We start by working in the regime where $m_N > m_\phi$, which is the most relevant for the displaced vertex searches in \cref{sec:Results}. We then discuss the opposite case, $m_\phi > m_N$. Using the benchmarks in \cref{eq:benchmarks}, the numerical results of this section are summarised in \cref{fig:lifetimes}.

\paragraph{$m_N > m_\phi$}: In the effective theory of \cref{eq:EffectiveNTP}, the pseudoscalar state has only one decay channel into two photons, \cref{eq:Photons:Effective}, and has a rate given by
\beq\label{eq:phitoggdecayrate}
\Gamma(\phi\to\g\g)\equiv \Gamma_{\phi\g\g} = \frac{g_\g^2 m_\phi^3}{64\pi}\,,
\eeq 

In the ISS model of \cref{sec:NTPfromISS}, a direct coupling to light neutrinos is also allowed, leading to the decay rate 
\beq\label{eq:PhiToNuNu}
\Gamma(\phi\to\nu\nu)\equiv \Gamma_{\phi\nu\nu} = \frac{\vert c_\nu\vert ^2 m_\phi}{32\pi} = \frac{\UN^4 m_\phi}{64\pi}\,.
\eeq
Note that $c_\nu$ is proportional to the mixing angle squared, \cref{eq:couplingLight}; in the small mixing region, we thus expect the branching ratio to be dominated by the photon decay, while in the large mixing region, the rate of invisible decays increases. Namely, we have
\beq\label{eq:PhiRateRatio}    
\frac{\Gamma_{\phi\nu\nu}}{\Gamma_{\phi\gamma\gamma}} = \frac{\UN^4}{g_{\gamma}^2 m_\phi^2} \sim 4 \times 10^{10}\times\UN^4 \times\left( \frac{g_\gamma}{10^{-4}~\mathrm{GeV}^{-1}} \right)^{-2} \times \left( \frac{m_\phi}{50~\mathrm{MeV}} \right)^{-2}\,,
\eeq
which indicates that the visible decay dominates the branching ratio for mixing angles $\UN^2\lesssim10^{-5}$. Here we are using the benchmark value of $g_\g$ considered in the analysis of \cref{sec:Results}. In this regime, the proper decay length of $\phi$ is
\beq
c\tau_{\phi\gamma\gamma}\sim 3\times10^{-2}~{\rm m}\times \lp \frac{50~{\rm MeV}}{m_\phi} \rp^3 \times \lp \frac{10^{-4}~{\rm GeV}^{-1}}{g_\g}\rp^2\,,
\eeq
that is, it will decay to photons after propagating for a few centimeters, independently of $U_N$; this displacement can be typically resolved by present and future detectors. Additional decay channels of $\phi$ into visible states can arise from propagating an off-shell $N$, which then decays to light SM fermions, as discussed below in \cref{eq:WeakDecay:Approx}; these processes are however 4-body decays and are suppressed by the weak Fermi coupling, $G_F$, thus we neglect them.

The HNL state decays to a light neutrino and a $\phi$ through the same transition lagrangian in \cref{eq:Transition:Effective}, with the rate given by 
\beq\label{eq:NtoPhiDecay:TwoBody}
\Gamma(N\to\nu\phi) \equiv \Gamma_{N\phi} =  \lambda_T^2\frac{m_N}{32\pi}\left(1-\frac{m_\phi^2}{m_N^2}\right)^2 = \UN^2\frac{m_N}{64\pi}\left(1-\frac{m_\phi^2}{m_N^2}\right)^2\,,
\eeq
where in the last equality we used \cref{eq:lambdaTtoMixing}. The proper decay length thus is
\beq
c\tau_{N\phi}\simeq 4\times10^{-8}~{\rm m}\times\lp \frac{10^{-6}}{\UN^2} \rp \lp \frac{{\rm GeV}}{m_N} \rp\,,
\eeq
for $m_\phi\ll m_N$. The NTP lagrangian thus introduces a new fast decay channel for the heavy neutrino. 

In the ISS model, we need to consider also decays induced by the weak current, \cref{eq:WeakMixing}. For $m_N\lesssim$ GeV, the main channels consist of decay to three leptons, or decay into a lepton and an hadron; for larger $m_N$, multi-meson decay rates start to dominate, which can be computed as the rate of $N$ decaying into a lepton and two quarks, times QCD loop corrections~\cite{Bondarenko:2018ptm}. Roughly, we have
\begin{align}\label{eq:WeakDecay:Approx}
    \Gamma(N\to3f) &\sim N_f\frac{G_F^2\UN^2}{192\pi^3} m_N^5\,, \\
    \Gamma(N\to\ell/\nu+{\rm h}^\pm/{\rm h}^0) &\sim \frac{G_F^2 f_h^2\UN^2}{32\pi} m_N^3\,,\label{eq:WeakDecay:Approx2}
\end{align}
where $N_f = 1, N_c,N_c|V_{ij}|^2$ if the three fermions in the final state are three leptons, a neutrino and two quarks (neutral current), a charged lepton and two quarks (charged current), respectively; here, $N_c = 3$ is the number of colors, and $V_{ij}$ the relevant CKM matrix element. The decay into a meson is instead proportional to the respective meson decay constant, $f_h$. 

As these are all weak interaction decays, the rates are all suppressed by $G_F^2$. It is then instructive to compare the lifetime associated with weak decays with that associated with the new channel. At $m_N\lesssim$ GeV, the two widths in \cref{eq:WeakDecay:Approx,eq:WeakDecay:Approx2} are approximately equal, and comprise almost the entire width of $N$; for the indicative scalings in this section, we then compute the total weak width, $\Gamma_W$, as twice the decay to three leptons. With this, we have
\beq\label{eq:DecayLengthWeak}
c\tau_W\sim 10^3~{\rm m}\times \lp \frac{10^{-6}}{\UN^2} \rp \lp \frac{{\rm GeV}}{m_N} \rp^5\,.
\eeq
The proper decay length through weak interactions is thus many orders of magnitude larger than the NTP-induced one in \cref{eq:NtoPhiDecay:TwoBody}. This is a crucial difference, as neutrino beam experiments rely on HNL produced far away, typically at the proton-on-target scattering stage, while the reach of collider experiments is limited by the long lifetimes; the NTP leads instead to a very fast chain of decays, allowing production and decay inside the detector volumes. We elaborate on this point in \cref{sec:Results}. 

\paragraph{$m_\phi > m_N$:} 
In this regime, the $N\to\nu\phi$ decay is kinematically forbidden, while the channel $\phi\to\nu N$ opens up. The latter has a rate
\beq
\Gamma(\phi\to\nu N)\equiv \Gamma_{\phi\nu N} = \frac{\lambda_T^2 m_\phi}{64\pi}\lp 1 - \frac{m_N^2}{m_\phi^2} \rp^2 = \frac{\UN^2 m_\phi}{128\pi}\lp 1 - \frac{m_N^2}{m_\phi^2} \rp^2\,.
\eeq
Unlike the decay to two light neutrinos, this rate scales as $\UN^2$; the ratio in \cref{eq:PhiRateRatio} changes accordingly, meaning that in this regime the visible decay of $\phi$ is relevant only for very small values of the mixing angle, $\UN^2\lesssim10^{-10}$. It also follows that in neutrino-mediated processes, $\phi$ can be produced only through an off-shell light or heavy neutrino, as computed schematically in \cref{eq:splitting}.

The two-body decay $N\to\nu\phi$ is now forbidden, and a decay to visible states needs an off-shell $\phi$, leading to a three-body process. The rates in this case are 
\begin{align}\label{eq:NdecayRates:Case2}
    \Gamma(N\to\nu+\g\g)\equiv  \Gamma_{N\g\g} &= \frac{\lambda_T^2 g_\gamma^2 }{192\pi^3} \frac{m_N^7}{16m_\phi^4} =  \frac{ \UN^2 g_\gamma^2 }{384\pi^3} \frac{m_N^7}{16m_\phi^4} \,, \\ 
    \Gamma(N\to\nu+\nu\nu)\equiv  \Gamma_{N\nu\nu} &= \frac{\lambda_T^2\vert c_\nu\vert ^2}{192\pi^3} \frac{m_N^5}{m_\phi^4} = \frac{\UN^6}{768\pi^3} \frac{m_N^5}{m_\phi^4}\,, 
\end{align}
and their ratio is
\begin{equation}
    \frac{\Gamma_{N\g\g}}{\Gamma_{N\nu\nu}}= \frac{g_\g^2m_N^2}{8\UN^4}\sim10^3\times \lp \frac{g_\g}{10^{-4}~{\rm GeV}^{-1}}\rp^2 \times \lp \frac{m_N}{\rm GeV}\rp^2 \times \lp\frac{10^{-6}}{\UN^2}\rp^2\,,
\end{equation}
that is, the two rates are comparable for $\UN\sim10^{-2}$, and the two photons decay dominates for smaller mixing angles.

However, the lifetime is much longer than in the previous scenario. In the effective NTP scenario, we have
\beq
c\tau_{N\g\g} \sim 2\times10^7~{\rm m} \times \lp \frac{10^{-3}}{\lambda_T} \rp^2 \times \lp \frac{10^{-4}~{\rm GeV}^{-1}}{g_\g}\rp^2\times \lp \frac{\rm GeV}{m_N} \rp^7\times \lp \frac{m_\phi}{10~{\rm GeV}} \rp^4\,.
\eeq
Comparing the same lifetime in the ISS model with the weak induced lifetime, we have
\beq\label{eq:Case2:NTPoverMixing}
    \frac{\Gamma_{N\g\g}}{\Gamma_W}\sim 10^{-4}\times \lp \frac{g_\g}{10^{-4}~{\rm GeV}^{-1}}\rp^2 \times \lp \frac{m_N}{\rm GeV}\rp^2  \times \lp \frac{10~{\rm GeV}}{m_\phi}\rp^4 \,.
\eeq
For heavy enough $N$ and $\phi$ the decay through weak currents is the fastest channel; however the large inverse power scaling with $m_\phi$ steeply changes this ratio across the mass range, and the photon channel can become competitive or dominant over the weak one. As an example, for $m_\phi = 50$ MeV and $m_N = 10$ MeV, this ratio becomes $\Gamma_{N\g\g}/\Gamma_W\sim20$.

\begin{table*}[t]
\centering
\renewcommand{\arraystretch}{1.2}
\begin{tabular}{lccc}
\hline\hline
Probe / signature & Standard HNL & ISS real. &  Effective NTP
\\
\hline
\hline
$M\to \ell +N$ & bump hunt & bump hunt& inclusive rate\\

Inclusive (NC/CC, $\Gamma_X$) & $\times$ & $\checkmark$ & $\checkmark$ \\
{\rm BBN} & $\checkmark$ & $\checkmark$ & $\checkmark$ \\
\hline
$M\to X\ell+$DV & $\checkmark$ & $\checkmark$ & $\checkmark$

\\

$W\to \ell +$DV, $Z\to \nu+ $DV & $\checkmark$ & $\checkmark$ & $\checkmark$ \\

Beam dump downstream searches & $\checkmark$ & $\times$ &$\times$ \\

DV from production in $\nu$ detectors & $\times$& $\checkmark$& $\checkmark$
\\

\hline\hline
\end{tabular}
\caption{Schematic comparison of the main experimental signatures described in \cref{sec:Results}, for the standard HNL scenarios, the ISS realization of the NTP, and the effective NTP model.}
\label{tab:HNL_NTP_signatures}
\end{table*}

%
\section{Experimental searches}
\label{sec:Results}
%

We present here the phenomenological analysis and the resulting regions of parameter space that can be probed at present and future experiments. 

We fix the benchmark values 
\beq\label{eq:benchmarks}
m_\phi=50~{\rm MeV}\,, \qquad g_\g=10^{-4}~{\rm GeV}^{-1}\,,
\eeq
for the pseudoscalar mass and its coupling to photons, respectively. These values are motivated by previous studies in Refs.~\cite{Bansal:2022zpi,Carey:2025oxt}; the $\phi$ state needs to be above the MeV scale to avoid strong astrophysical and cosmological constraints\footnote{The 'SN-diffuse $\g$' line shown in Fig.~7 of \cite{Carey:2025oxt} was wrongly reported, and its exclusion region largely overestimated. An erratum is in preparation.}, while collider searches exclude light ALP couplings larger than $g_\g\simeq10^{-4}~{\rm GeV}^{-1}$. We then consider two setups, which follow along the discussion of \cref{sec:Model}: a pure NTP lagrangian, \cref{eq:EffectiveNTP}, and an ISS inspired one, \cref{eq:ISS}. In the former case, we take $m_N$ and $\lambda_T$ as free parameters, while in the latter we take $m_N$ and $|U_N|$; for simplicity, we assume that in both models the heavy neutrino $N$ only couples and mixes with the muon neutrino flavor state. In the ISS model, we highlight in green the preferred region to reproduce the neutrino mass, as determined by \cref{eq:numassregion}, while for the NTP we show with the same color the region where loop corrections to the neutrino mass do not exceed the limits on the observed one, see \cref{eq:NeutrinoMassLoopBound}. Our results are summarized in \cref{fig:Results}. In \cref{sec:otherbenchmarks}, we briefly discuss how the phenomenology changes for two different benchmarks: a heavy (pseudo)scalar mediator, $m_\phi=500~{\rm MeV}$, and a fully invisible one, $g_\g=0$.

\begin{figure}[t!]
    \centering
    \includegraphics[width=0.49\linewidth]{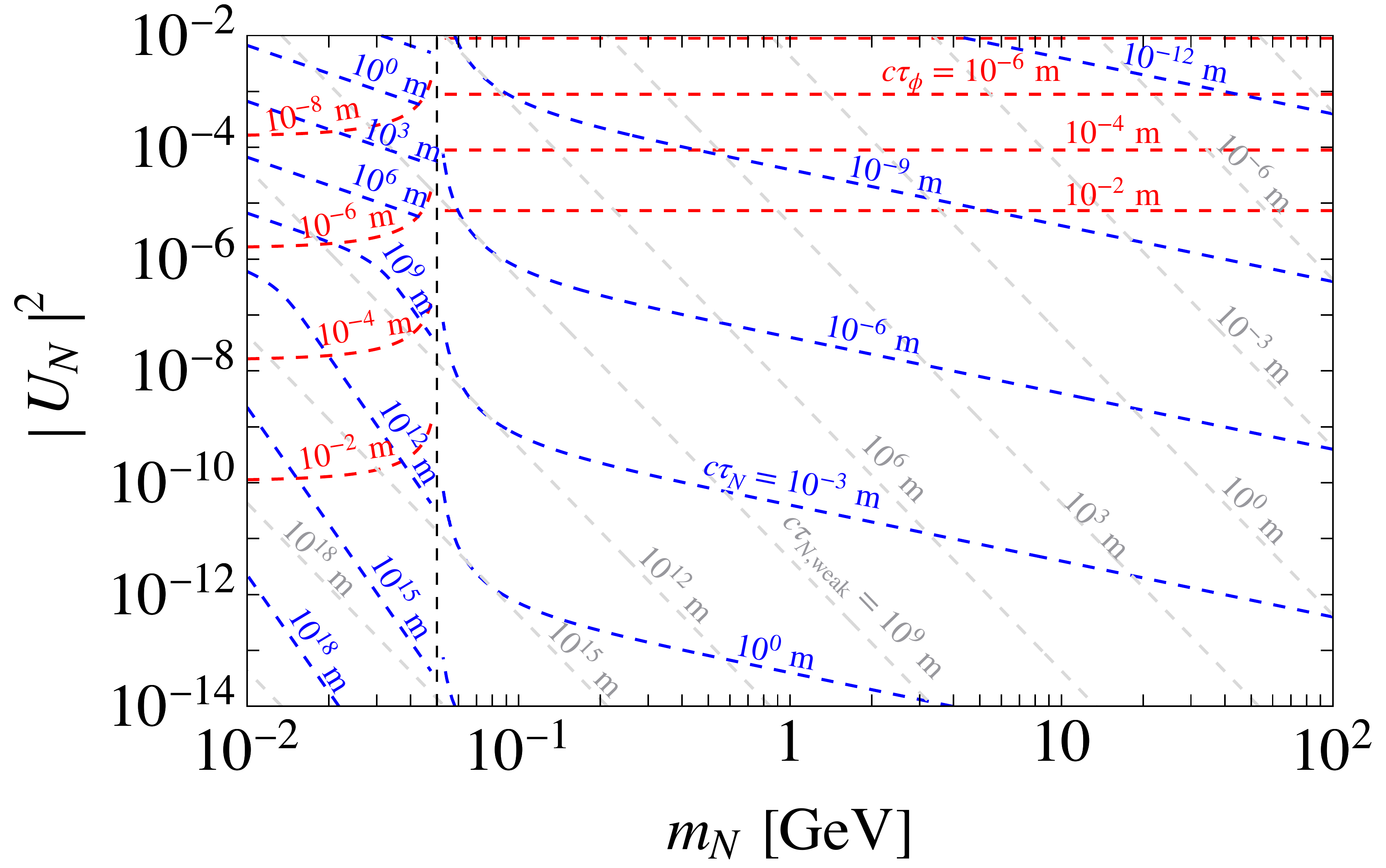}
    \includegraphics[width=0.49\linewidth]{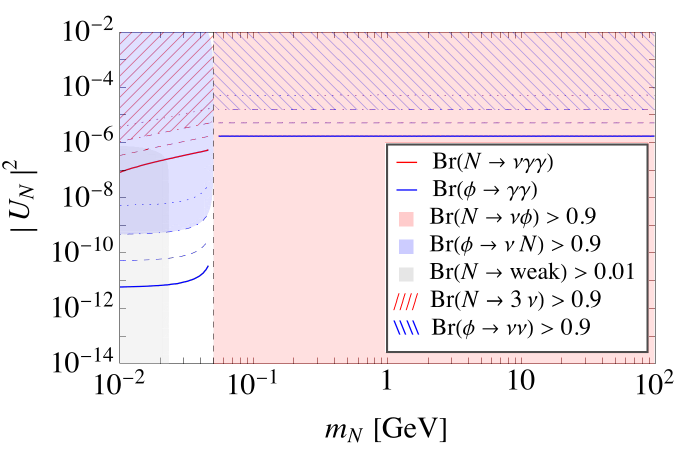}
    \caption{Proper decay lengths and branching fractions of the unstable states in the ISS model, in the $(m_N,|U_N|^2)$ plane. In both panels, the vertical black dashed line marks the kinematic threshold $m_N=m_\phi=50$ MeV. \textbf{Left:} Isocontours of proper decay length for $\phi$ (red) and $N$ (blue). Gray dashed contours show the $N$ decay length considering only weak interactions. \textbf{Right:} Isocontours of ${\rm Br}(N\to\nu\gamma\gamma)$ (red) and ${\rm Br}(\phi\to\gamma\gamma)$ (blue). Shaded regions indicate where ${\rm Br}(N\to\nu\phi)>0.9$ (red), ${\rm Br}(\phi\to\nu N)>0.9$ (blue), and where the total branching fraction of weak-induced $N$ decays exceeds $1\%$ (gray). Hatched regions correspond to ${\rm Br}(N\to3\nu)>0.9$ (red) and ${\rm Br}(\phi\to\nu\nu)>0.9$ (blue).}
    \label{fig:lifetimes}
\end{figure}

\begin{figure}[t!]
    \centering
    \includegraphics[width=0.75\linewidth]{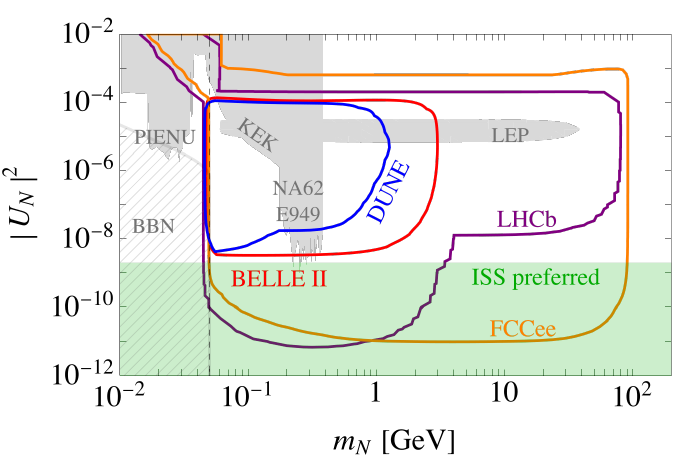}
    \includegraphics[width=0.75\linewidth]{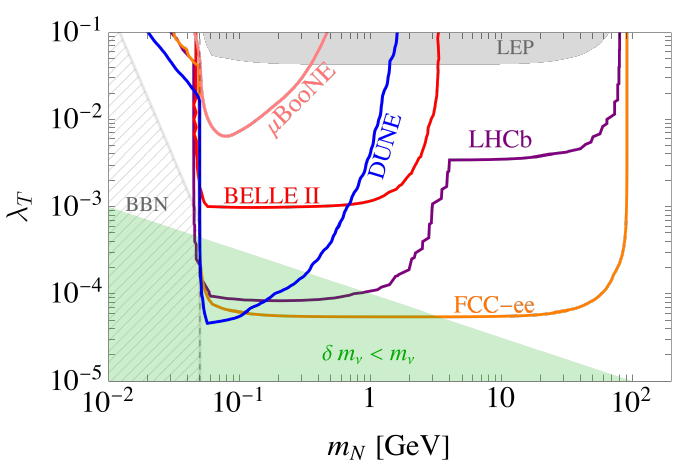}
    \caption{Projected sensitivity for the ISS model in the $(m_N,|U_N|^2)$ plane ({\bf top}) and for the NTP one in the $(m_N,\lambda_T)$ plane ({\bf bottom}), where the vertical black dashed line marks the kinematic threshold $m_N=m_\phi=50$ MeV. In the former, the green region indicates the parameter space where the ISS model can reproduce the neutrino mass, \cref{eq:numassregion}; in the latter, it shows the region where loop corrections to the neutrino mass do not exceed the limits on the observed one, \cref{eq:NeutrinoMassLoopBound}. In both cases, we show in gray existing limits, invisible signals from PIENU, KEK and NA62/E949 (\cref{sec:InvisibleSignals}) and displaced from LEP (\cref{sec:epemColl}); colored lines show the strongest projected bound from displaced vertex searches discussed in \cref{sec:VisibleSignals}: DUNE (blue), BELLE-II (red), LHCb (purple), and FCC-ee (orange). The hatched region indicates exclusions from BBN.
   }
    \label{fig:Results}
\end{figure}

As pointed out already in \cref{sec:Decay}, the main phenomenological consequence of the NTP lagrangian is the additional decay channel of the heavy neutrino state. Taking the benchmark values of \cref{eq:benchmarks}, the proper lifetimes of $N$ and $\phi$ in the ISS model are shown in the left plot of \cref{fig:lifetimes}; the branching ratio into invisible and visible final states is shown in the right plot of \cref{fig:lifetimes}. By combining the lifetime and branching fraction information, we can understand which type of signal is expected in the two mass regions of interest. In the $m_N<m_\phi$ region, the new states are either long-lived or decay predominantly into neutrinos; we describe these \emph{invisible} signals in \cref{sec:InvisibleSignals}. Conversely, in the $m_N>m_\phi$ regime both $N$ and $\phi$ are short-lived and the dominant final state contains multiple photons; here we expect new \emph{displaced} signals, which we explore in \cref{sec:VisibleSignals}. The types of signal explored in this work are summarised in \cref{tab:HNL_NTP_signatures}.

Finally, note that we are presenting \cref{fig:lifetimes} for the ISS model. The only sizeable difference with the effective NTP model is the invisible decay channel $\phi\to\nu\nu$, \cref{eq:PhiToNuNu}, which modifies the branching fraction into visible final states for large mixing values; this is discussed further in \cref{sec:VisibleSignals}. The weak mixing induced decays are negligible for the phenomenology of interest in this benchmark (see the gray region in \cref{fig:lifetimes}); thus, the signal description is valid also for the effective NTP search, by using the scaling in \cref{eq:lambdaTtoMixing}. They are instead relevant in the two benchmarks discussed in \cref{sec:otherbenchmarks}.

\subsection{Invisible signals}
\label{sec:InvisibleSignals}

Below the $m_\phi$ threshold, the $N$ lifetime is dictated by the three-body decay rates in \cref{eq:NdecayRates:Case2}, with a decay length above the kilometer scale already at large mixing values. The branching fraction is dominated by the invisible decay $N\to3\nu$ for $\UN^2\gtrsim10^{-6}$, while the visible decay $N\to\nu\g\g$ takes over only at $\UN^2\lesssim10^{-6}$, where, however, $N$ is effectively stable on Earth distance scales. Conversely, $\phi$ is short-lived, but decays invisibly for $\UN^2\gtrsim10^{-9}$, predominantly to $N\nu$. We thus expect this region to be dominated by \emph{invisible} signatures; we describe them in the following.

\subsubsection{Kinematic features in the final state}
\label{sec:KinematicFinalState}
New states can be searched for inclusively by looking for kinematic features in the momentum or energy spectra of Standard Model particles produced alongside them. A clear example is the semileptonic meson decay $M\to\ell+X$, where $X$ represents the inclusive final state. While the limit from these decays is in general model-independent, the search strategy depends on the combined final state. If the SM side is visible, one can reconstruct the energy and invariant mass spectrum, and search for distortions; in the simplest case of a 2-body decay, this distortion would appear as a shifted peak in the SM spectrum. If part of the SM is not visible, then one can measure modifications of the inclusive total rate.

In \cref{fig:Results} we include limits from PIENU~\cite{PIENU:2019usb}, which searched for a peak in the muon momentum spectrum from $\pi^+ \to \ell X$, and from NA62 \cite{NA62:2021bji}, KEK \cite{Hayano:1982wu}, and E949 \cite{E949:2014gsn}, which performed analogous searches for peaks in the muon momentum from $K^+ \to \ell X$. These are shown as gray regions in the upper plot of \cref{fig:Results}. Note that similar regions are absent in the lower plot, because in the effective NTP scenario, the final state is $X=N+\phi$; the bump hunt search does not apply, and it is replaced by a shape analysis of the lepton spectra, which reduces the reach.

An additional limit can come from searches at BABAR~\cite{BaBar:2022cqj} for kinematic distortions in the tau three-prong decay, $\tau^\pm\to\pi^\pm\pi^\mp\pi^\pm X$, if one relaxes the working assumption that $N$ mixes only with muon neutrinos. While it only limits $\UN^2\sim10^{-2}$ for $m_N\lesssim300$ MeV, the bound becomes more important closer to threshold, reaching $\UN^2\sim10^{-5}$ for $m_N\sim1$ GeV. However, a caveat to all the aforementioned searches is that the final state $X$ is assumed to remain invisible in the detector; a prompt decay, especially in the $m_N>m_\phi$ region, can occur inside the apparatus, and modify the observed signal. In particular, the BABAR analysis explicitly assumed that $N$ did not decay inside the detector and imposed vetoes on additional neutral calorimeter activity. A prompt decay could consequently reduce the signal acceptance, so the published limit applies directly only when $N$ is effectively invisible on the scale of the BABAR detector.

\subsubsection{Indirect bounds}
\label{sec:IndirectBounds}

\paragraph{Inclusive neutrino NC/CC:}
The NTP vertex represents a genuinely new interaction and introduces an additional physical final state. Unlike the standard HNL scenario, the inclusive rate of SM NC or CC neutrino interactions can be modified\footnote{An analogous enhancement has been proposed in Refs.~\cite{Borrello:2025hal,Borrello:2026bfr} in the context of strongly coupled dark sectors.}. Using the estimate in \cref{eq:neutrinothreebody}, we have~\cite{Davidson:2003ha} 
\beq\label{eq:ncccmodification}
R_\nu \equiv \frac{\sigma_{\rm NC}}{\sigma_{\rm CC}}= R_\nu^{\rm SM}\left(1+\Delta R_\nu\right)\,,
\qquad
\Delta R_\nu\sim\frac{|\lambda_T|^2}{16\pi^2}\,.
\eeq
The last step is simplified by neglecting CC modifications, which must proceed through the $N$ light-heavy mixing, if present, giving
\beq
\frac{\sigma_{\rm CC+\phi}}{\sigma_{\rm CC}^{\rm SM}}
\sim
|U_N|^2
\frac{|\lambda_T|^2}{16\pi^2}L\,.
\eeq
The current precision on $\Delta R_\nu\sim 10^{-3}$ comes from NuTeV~\cite{NuTeV:2001whx}, and allows to exclude $\lambda_T\gtrsim 10^{-1}$; thus, it does not appear in \cref{fig:Results}. Future improvements in the statistics of neutrino-scattering measurements, in particular at DUNE~\cite{DUNE:2021tad}, may extend the sensitivity to $\lambda_T\lesssim 10^{-2}$, as shown in \cref{sec:otherbenchmarks}. In the limit $m_N\to 0$, inclusive processes are unaffected by standard HNL, which acts as a unitary change of basis. This is strictly true only in the massless limit as ${\cal O}(m_N^2/s)$ corrections arise from phase space corrections; these corrections are relevant only at kinematic closure and are not considered further in this work.

\paragraph{SM branching ratios:} The emission of $\phi$, whose parametric size can be estimated from \cref{eq:neutrinothreebody,eq:decaywidthmodification}, can also modify the inclusive decay rates of SM particles. In particular, a decay $X\to X'+{\rm anything}$ receives an additional contribution from the emission of an $N\phi$ pair through an off-shell neutrino, $X\to X'N\phi$. This genuinely new channel contributes to the inclusive decay width, which can be parametrically written as
\beq\label{eq:WidthShift}
\Gamma_X=\Gamma_X^{\rm SM}(1+\Delta\Gamma_X)\,,\qquad \Delta\Gamma_X\sim \frac{|\lambda_T|^2}{16\pi^2}\,,
\eeq
where $\Gamma_X^{\rm SM}$ is the inclusive decay width in the SM.
Relevant observables include the invisible $Z$ width, measured at LEP~\cite{ALEPH:2005ab} with a relative precision of a few per mille, with substantial improvements expected at FCC-ee~\cite{Bolton:2025tqw,AlcarazMaestre:2021ssl}, as well as the leptonic branching fractions of $\tau\to\mu\bar\nu_\mu\nu_\tau$~\cite{ParticleDataGroup:2026mpi,ALEPH:2005qgp,Belle-II:2024vvr} and $K\to\mu\bar\nu_\mu$~\cite{ParticleDataGroup:2026mpi}. As shown in \cref{sec:otherbenchmarks}, current measurements typically yield bounds of ${\cal O}(1)$ on $\lambda_T$, while future precision measurements may improve the sensitivity. As for NC/CC, up to phase space corrections which we neglect in \cref{tab:HNL_NTP_signatures}, enhancements of inclusive processes are absent in standard HNL scenarios.

\paragraph{Radiative correction to neutrino mass:} 
For a Majorana $N$, the $N$--$\phi$ loop generates a contribution of naive size $\delta m_\nu\sim m_N\lambda_T^2/(16\pi^2)$. Requiring $\delta m_\nu<m_\nu$ gives the condition
\begin{equation}\label{eq:NeutrinoMassLoopBound}
|\lambda_T|\lesssim 10^{-4} \left(\frac{1~{\rm GeV}}{m_N}\right)^{1/2}\,.
\end{equation}
Note that this limit strictly depends on the Majorana nature of $N$; it is absent if $N$ is a Dirac neutrino, as it requires a breaking of the lepton number. Similarly, in the ISS realization the only mass correction is proportional to the small lepton number breaking parameter, $f_\phi$, and is thus negligible. For the NTP model, we show this preferred region of the parameter as a green region in the bottom panel of \cref{fig:Results}, stressing that this boundary applies only to the Majorana realization.

\paragraph{Cosmology:} 
We include possible limits from cosmological observables among the invisible signatures. The effect of the light pseudoscalar $\phi$ couplings to neutrinos and photons has been analysed already in Ref.~\cite{Bansal:2022zpi}; these limits are the main reason to consider $m_\phi \gtrsim10$ MeV, as above this value cosmological limits evaporate.

For $N$, we consider a simplified approach and show the lifetime limit from BBN, computed in Ref.~\cite{Ruchayskiy:2012si}. The full computation can become cumbersome; at large mixing angles, $\UN^2\gtrsim10^{-5}$, the decay is dominated by the $3\nu$ mode, with a smaller contribution from the $\nu\g\g$ mode, which then becomes dominant at lower mixings. These two decay modes can have competing effects on the value of $N_{\rm eff}$, and the thermal history of the $N$ sector needs to be tracked in detail. We then lightly hash the region in \cref{fig:Results}, corresponding to $\tau_N\gtrsim 0.1$ s, emphasizing that it should be interpreted only as an indicative BBN-sensitive region rather than a robust exclusion.

\subsection{Visible signatures}
\label{sec:VisibleSignals}

Above the $m_\phi$ threshold, the $N\to\nu\phi$ channel is allowed, dramatically shortening the $N$ lifetime, as seen in \cref{fig:lifetimes}; a decay length of the order of the millimeter is reached only at $\UN^2\sim10^{-10}$. The lifetime of $\phi$ is relatively unchanged, as the $N\nu$ channel is closed, but the $\nu\nu$ and $\g\g$ remain. The latter channels then determine the rate of visible signals. Note that the $\nu\nu$ channel is absent in the effective NTP lagrangian, and thus the decay of the chain $N\to\nu\phi,~\phi\to\g\g$ is visible. 

A direct consequence of shortening the $N$ lifetime is that the NTP model cannot be searched for downstream from beam dumps. These searches rely both on the large production of $N$ from proton beam dumped on targets, and on the long lifetime of $N$ from weak decays; the decay products can then be observed in a distant downstream detector. Examples of such experiments are NuTeV~\cite{NuTeV:1999kej}, CHARM~\cite{CHARM:1983ayi,CHARM:1985nku,Boiarska:2021yho}, and PS191~\cite{Bernardi:1985ny,Bernardi:1987ek}, whose typical constraints on standard HNL vanish in the presence of the $N\to\nu\phi$ decay channel. 

We then focus here on visible signals, where the $N$ and $\phi$ are both produced and decay in the instrumented detector volume. Furthermore, the initial interaction that produces $N$ often contains additional visible states, or has a known interaction point (IP). Thus, to avoid most of the experimental backgrounds and have reliable projections for future experiments, we focus on processes with a visible displaced vertex (DV). 

In the approximation that $N$ and $\phi$ are collinear, the position of the $\phi\to\gamma\gamma$ decay vertex with respect to the production point of $N$ is determined by the sum of the distances traveled by the two particles,
\begin{equation}
    \ell_{\rm DV}=\ell_N+\ell_\phi\,.
\end{equation}
For a given event, they are exponentially distributed with decay lengths in the laboratory frame
\begin{equation}
    L_N\equiv \beta_N\gamma_N c\tau_N=\frac{|\mathbf p_N|}{m_N}c\tau_N\,,\qquad L_\phi\equiv\beta_\phi\gamma_\phi c\tau_\phi=\frac{|\mathbf p_\phi|}{m_\phi}c\tau_\phi\,.
\end{equation}
The probability that the $\phi\to\gamma\gamma$ vertex is displaced by at least $L_{\min}$, while occurring before a maximal detector length $L_{\rm det}$, is therefore obtained by convoluting the two exponential
decay distributions,
\begin{equation}
\begin{split}
    P_{\rm displ} &=\int_0^{L_{\rm det}}\frac{d\ell_N}{L_N}\,e^{-\ell_N/L_N}\left[e^{-\max(0,L_{\min}-\ell_N)/L_\phi} - e^{-(L_{\rm det}-\ell_N)/L_\phi}\right]\\ &=\frac{L_N e^{-L_{\min}/L_N} - L_\phi e^{-L_{\min}/L_\phi}}{L_N-L_\phi} -\frac{ L_N e^{-L_{\rm det}/L_N} -L_\phi e^{-L_{\rm det}/L_\phi}}{L_N-L_\phi}\,,
    \label{eq:PDVcascade}
\end{split}
\end{equation}
for $L_N\neq L_\phi$. In the degenerate limit $L_N=L_\phi\equiv L$, this reduces to
\begin{equation}
    P_{\rm displ} =e^{-L_{\min}/L}\left(1+\frac{L_{\min}}{L}\right)- e^{-L_{\rm det}/L}\left(1+\frac{L_{\rm det}}{L}\right).
\end{equation}
Since the decay of $N$ is effectively prompt in the regime considered, the probability of observing a DV is well approximated by the probability for $\phi$ to decay between the positions $L_{\rm min}$ and $L_{\rm max}$,
\begin{equation}
    P_{\rm displ}^\phi(L_{\rm min},L_{\rm max})
    =
    \exp\left(
    -\frac{L_{\rm min}}{L_\phi}
    \right)
    -
    \exp\left(
    -\frac{L_{\rm max}}{L_\phi}
    \right)\,.
    \label{eq:Pphi}
\end{equation}
This well-known result can be reproduced by \cref{eq:PDVcascade} in the prompt $N$ limit, $L_N\to0$.

Finally, note that in the $m_N>m_\phi$ regime, the $\phi$ is boosted, with a Lorentz factor that depends on the energy distribution of the parent $N$.  

Neglecting angular correlations, its average boost is
\begin{equation}
    \left\langle\gamma_\phi\right\rangle
    \simeq
    \gamma_N\frac{E_\phi^\star}{m_\phi},
    \qquad
    E_\phi^\star =
    \frac{m_N^2+m_\phi^2}{2m_N},
\end{equation}
which, in the limit $m_\phi\ll m_N$, reduces to
\begin{equation}
    \left\langle\gamma_\phi\right\rangle
    \simeq
    \frac{E_N}{2m_\phi}.
\end{equation}
For a relativistic $\phi$, its mean laboratory decay length is therefore approximately
\begin{equation}
    \left\langle L_\phi^{\rm lab}\right\rangle
    \simeq
    0.6~{\rm m}
    \left(\frac{E_N}{2~{\rm GeV}}\right)
    \left(\frac{50~{\rm MeV}}{m_\phi}\right)^4
    \left(\frac{10^{-4}~{\rm GeV}^{-1}}{g_\gamma}\right)^2 \,.
    \label{eq:boostedLphi}
\end{equation}
In the mass region considered, the DV rate depends also on the visible branching ratio
\beq
{\rm Br}({\rm vis})\approx  {\rm Br}(N\to\nu\phi)\,
    {\rm Br}(\phi\to\gamma\gamma)\,.
\eeq
In the effective NTP scenario, ${\rm Br}({\rm vis}) = 1$ in the entire parameter space of interest; in the ISS scenario, ${\rm Br}(\phi\to\gamma\gamma)\gtrsim0.9$ for $\UN^2\lesssim10^{-6}$, while above this mixing the ${\rm Br}(\phi\to\nu\nu)$ is relevant, see \cref{fig:lifetimes}, reducing the rate of visible final states.

\subsubsection{DUNE}
\label{sec:DUNE}

Experiments with intense neutrino beams provide a unique environment to probe NTP interactions; the $N$ can be produced alongside a photon through a coherently enhanced $\phi$-induced electromagnetic interaction. The usual decay chain then leads to a DV in the detector. We consider here the DUNE Near Detector setup \cite{DUNE:2021tad}, composed of 64 tonnes of liquid Argon; the flux of $\nu_\mu$ at detector is taken from Ref.~\cite{DUNE:2021cuw}. A similar setup is presently used by MicroBooNE~\cite{MicroBooNE:2025ntu}, albeit with a less intense neutrino beam, which leads to negligible constraints.

The number of events can be computed as
\beq \label{eq:nuevtrate}
N_{\rm evt}=N_{\rm target}\times {\cal A}\times \int \dd E_\nu \dd Y~\phi(E_\nu)\frac{\dd\sigma}{\dd Y}\times P_{\rm displ}\times {\rm Br}({\rm vis})
\eeq
where $N_{\rm target}$ is the total number of nuclear targets in the detector, ${\cal A}$ is the surface of the detector transverse to the beam direction,  $\phi(E_\nu)$ is the neutrino flux per unit area, $\dd\sigma/\dd Y$ is the $N$ production differential cross-section, with $Y$ representing the set of independent variables (see e.g. \cref{eq:NTP:nucleus:xs}). 

\begin{figure}[t!]
    \centering
    \includegraphics[width=0.49\linewidth]{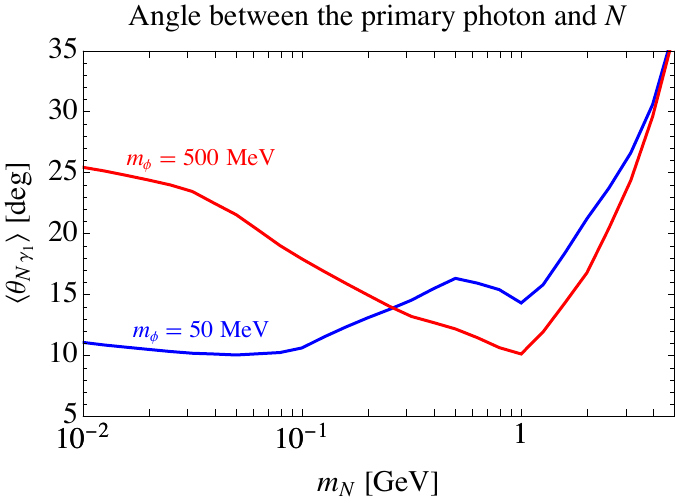}
    \includegraphics[width=0.49\linewidth]{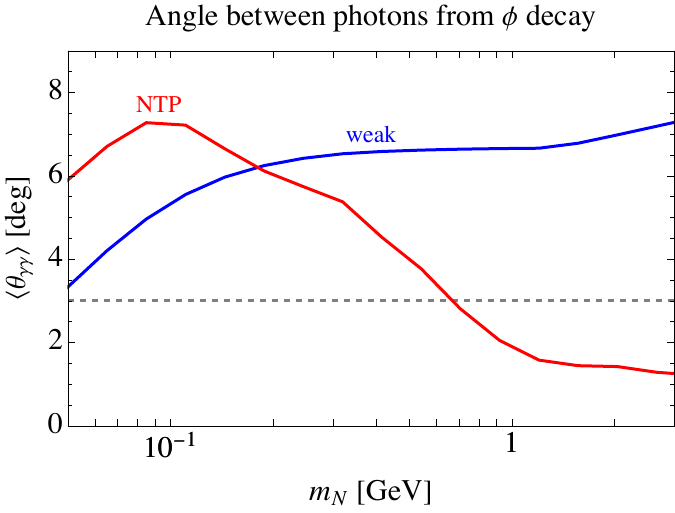}
    \includegraphics[width=0.49\linewidth]{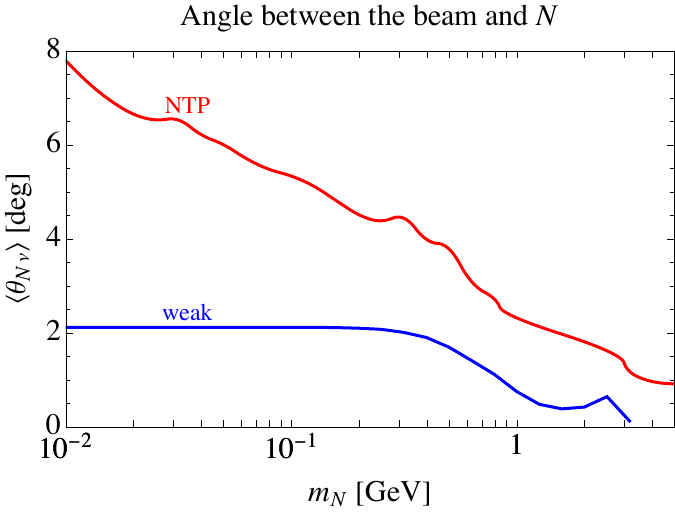}
\caption{Mean angular scales relevant for photon signatures from $N$ production and decay at DUNE, as a function of $m_N$. The angular distributions are averaged over the DUNE neutrino flux, weighted by the corresponding production cross sections. \textbf{Top left:} Mean angle between the HNL and the primary photon emitted in NTP production, \cref{fig:ntp}, for $m_\phi=50$ MeV (blue) and $500$ MeV (red). \textbf{Top right:} Mean opening angle between the two photons from $\phi\to\gamma\gamma$, for NTP-induced (red) and active--sterile mixing-induced (blue) production. The gray dashed line marks an angular separation of $3^\circ$. \textbf{Bottom:} Mean angle between the produced HNL and the incoming neutrino beam for NTP-induced (red) and weak (blue) up-scattering.}
    \label{fig:thetagammaN}
\end{figure}

The principal experimental challenge is the reconstruction of the photon decay products in the detector. In particular, we need the first photon to be clearly separated from the two displaced ones, and these two photons to be separated enough to reconstruct them. Approximately, two photons in the DUNE-ND can be separated if their angular aperture is $\theta_{\g\g}\gtrsim3\degree$.
These two angles are shown in the upper panel of \cref{fig:thetagammaN}, where we compute their average by weighting the angular distribution over the nominal DUNE-ND flux. In the upper left plot, we show the angle between the initial photon and $N$, $\theta_{N\g_1}$; this is always larger than $10\degree$ for both $m_\phi=50,500$ MeV, so the primary photon is unlikely to be collinear with the final two. In the upper right plot, we show the di-photon angular separation, $\theta_{\g\g}$. For NTP production, this is larger than $3\degree$ up to $m_N\sim1$ GeV, above which the boost from the $N$ decay leads to almost collinear photons; at this mass, also the cross section sharply drops (see \cref{fig:nuxsplot}), limiting the DUNE reach before the angle requirement is needed.

In the ISS scenario, it is also possible to produce $N$ through weak currents; the comparison of the two production cross sections is shown in \cref{fig:nuxsplot}. Unlike the NTP-induced coherent production, the weak process does not contain a prompt photon associated with the primary interaction vertex. The signal is therefore provided by an isolated photon pair, with a reconstructed direction that is separated from the beam axis. The latter angle, $\theta_{N\nu},$ is shown in the bottom panel of \cref{fig:thetagammaN}: it is $\sim2\degree$ across the mass range, and drops only at $m_N\sim1$ GeV, where anyway the cross section also drops (see \cref{fig:nuxsplot}). A relevant background arises from $\pi^0$ production, followed by $\pi^0\to\g\g$, which however can be efficiently suppressed by reconstructing the diphoton invariant mass and vetoing the region around $m_{\g\g}\simeq m_{\pi^0}$; this is possible as the two final photons are sufficiently separated to allow an independent reconstruction of their directions. 

\begin{figure}[t!]
    \centering
    \includegraphics[width=0.7\linewidth]{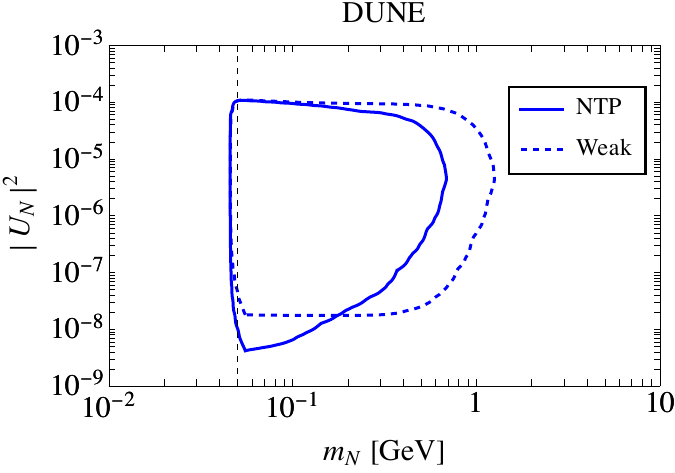}
    \caption{Projected sensitivity at DUNE in the $(m_N,|U_N|^2)$ plane, obtained by requiring three signal events. The solid blue contour corresponds to NTP production, while dashed blue contours show weak production through active--sterile mixing.}
    \label{fig:duneresume}
\end{figure}

The DUNE limits are then shown as a blue line in \cref{fig:Results}; in \cref{fig:duneresume} we explicitly break down the two contributions, showing the limits assuming only NTP (solid) and weak (dashed) production. A similar search can be performed at MicroBooNE~\cite{MicroBooNE:2021sov,MicroBooNE:2025ntu,MicroBooNE:sepangle}. In the pure-NTP scenario, the corresponding sensitivity is shown as a pink contour in \cref{fig:Results}. In the ISS benchmark, however, the parameter space within MicroBooNE's sensitivity is dominated by the invisible decay $\phi\to\nu\nu$, substantially reducing the observable photon signal and precluding significant constraints.

\subsubsection{LHCb}
\label{sec:LHCb}
\begin{figure}[t!]
    \centering
    \includegraphics[width=0.7\linewidth]{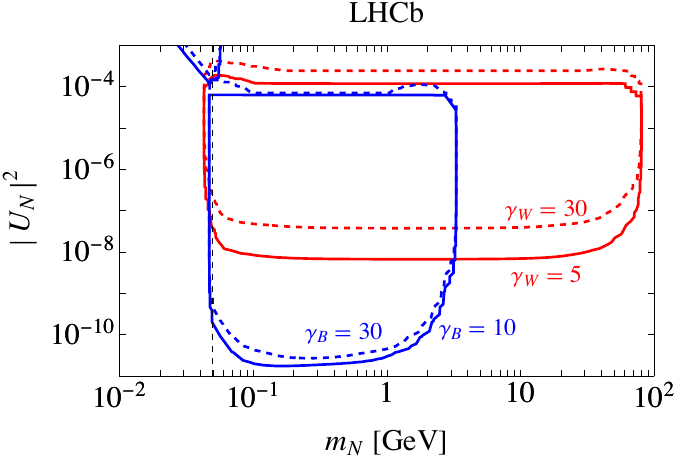}
    \caption{Projected sensitivity at LHCb in the $(m_N,|U_N|^2)$ plane, assuming an integrated luminosity of $300~{\rm fb}^{-1}$. Blue contours correspond to HNL production from $B$-meson decays, while red contours correspond to production from $W$-boson decays. Solid and dashed contours illustrate the dependence on the characteristic parent boost, with $\gamma_B=10,30$ and $\gamma_W=5,30$, respectively.}
    \label{fig:lhcbresume}
\end{figure}

The LHCb detector~\cite{LHCb:2008vvz,LHCb:2018roe,LHCb:2014osd,LHCb:2016inz,LHCb:2025ymr} can exploit the large LHC luminosity and its forward direction to look for displaced decays of both $B$ mesons and $W^\pm$ bosons. Once produced at the IP, these particles decay through the processes described in \cref{eq:MixingDecays,eq:NoMixingDecays}; the charged-lepton track can help reconstruct the primary vertex, while the di-photon final state is observed in the forward detector. The overall result is shown as a purple line in \cref{fig:Results}; the two cases are shown separately in \cref{fig:lhcbresume}.

In both cases, the number of expected events is 
\beq\label{eq:NeventsDispl:Bosons}
N_{\rm evt}^i = N_i\times{\rm Br}(i\to N+{\rm X})\times{\rm Br}_{\rm vis}\times P_{\rm displ}^\phi\times\epsilon_{\rm geo}\,,
\eeq
where $i=B,W$, $X$ collectively denotes any additional particles accompanying $N$ in the final state, and $N_i={\cal L}\times\sigma_{\rm prod}$ is the total number of parent particles produced.  The factor $\epsilon_{\rm geo}$ denotes the conditional probability that, once the parent is produced, the subsequent $N\to\nu\phi$, $\phi\to\gamma\gamma$ decay chain also remains within the instrumented angular region.

For the highly boosted forward production relevant at LHCb, the decay products are approximately collimated with the parent direction. We therefore encode the detector angular acceptance in $\epsilon_{\rm geo}$ by applying it to the parent particle, $i=B,W$, and assume that a parent produced within the detector acceptance yields a decay chain that remains inside the instrumented region. Notice that the boost dependence of the decay length is not included in $\epsilon_{\rm geo}$, but separately in $P_{\rm displ}^{\phi}$ through its decay length $L_\phi$. The latter is evaluated by requiring the displaced decay to occur downstream of the VELO and before the electromagnetic calorimeter, approximately within the fiducial interval $1~{\rm m}<L<12~{\rm m}$ from the IP.

Starting with the $B$-meson case, at the full integrated luminosity of ${\cal L}=300~{\rm fb}^{-1}$, we expect ${\cal O}(10^{13})$ $B$ mesons to be produced in the LHCb fiducial region; see, for instance, Ref.~\cite{LHCb:2016qpe} for a measurement of the $b$-hadron production cross section at LHCb. For simplicity in this work we focus on the benchmark channel
\begin{equation}
    B\to D^{(*)}\mu N\,,
\end{equation}
as it provides one of the largest production rates~\cite{Bondarenko:2018ptm}. Furthermore, the charged tracks from the $D^{(*)}$ decay, together with the prompt muon, can be used to reconstruct the $B$ decay vertex and trigger the event.  More inclusive channels, such as $B\to \mu N X$, can increase the signal yield, while leptonic $B_c\to\mu N$ decays can extend the sensitivity to somewhat larger masses. 

The parent $B$ meson kinematics at LHCb span a relatively broad distribution. Rather than performing an event-by-event convolution over the measured spectrum, we estimate the associated uncertainty by varying the characteristic boost over the representative range $\gamma_B=10$--$30$, motivated by the measured $B$-meson $p_T$ and rapidity distributions at LHCb~\cite{LHCb:2017vec}. As shown in \cref{fig:lhcbresume}, the resulting variation of the projected sensitivity is modest over most of the parameter space, indicating that the reach is not strongly dependent on the precise value adopted for the parent boost. The projected sensitivity from $B$ meson decays is comparable to the DUNE reach, \cref{fig:Results}, while extending to larger $m_N$ due to the higher kinematic reach of $B$ decays.

Even larger $N$ masses can be reached by searching for DVs in $W$ boson decays. With full luminosity we expect ${\cal O}(10^{10})$ $W$ bosons to be produced in the LHCb fiducial region; see, for instance, Ref.~\cite{LHCb:2026dan} for a recent measurement of forward $W$ production at LHCb. The $W$ boson kinematics are characterized by a broad forward distribution. We extract the corresponding energy and rapidity distributions from a {\tt MadGraph} simulation imposing the LHCb fiducial cuts, finding typical boosts of ${\cal O}(10)$, consistent with the distributions reported in Refs.~\cite{ATLAS:2025mlt,LHCb:2026dan}. As for the $B$-meson case, we assess the impact of the boost approximation by varying the representative value in the range $\gamma_W=5$--$30$, as shown in \cref{fig:lhcbresume}.

\subsubsection{Electron-positron colliders: Belle II, LEP and FCC-ee}
\label{sec:epemColl}

Searches of DVs at electron positron colliders provide a complementary probe of heavy neutrinos $N$, combining the clean environment from $e^+e^-$ scatterings with large statistics. Here we consider DVs in $B$ meson decays at Belle~\cite{Belle:2012iwr} and Belle II~\cite{Belle-II:2018jsg} and $Z$ boson decays at LEP~\cite{DELPHI:1996qcc,ALEPH:2005ab} and FCC-ee~\cite{FCC:2025lpp}. In each case, the expected number of displaced events follows again directly from \cref{eq:NeventsDispl:Bosons}, with $i=B,Z$ respectively; the search for DV proceeds then along the same lines as for LHCb. The main difference lies in the kinematics: $Z$ bosons are produced resonantly at rest at LEP and FCC-ee, while $B$ mesons receive a small boost, $(\gamma\beta)_B\simeq0.28$, due to the asymmetry in the electron and positron beam energies, which is however negligible for the purposes of this analysis. It follows that in each case the geometrical acceptance is equivalent to the solid angle coverage of the main detector, as the decay chain is isotropically distributed.

\begin{figure}[t!]
    \centering
    \includegraphics[width=0.7\linewidth]{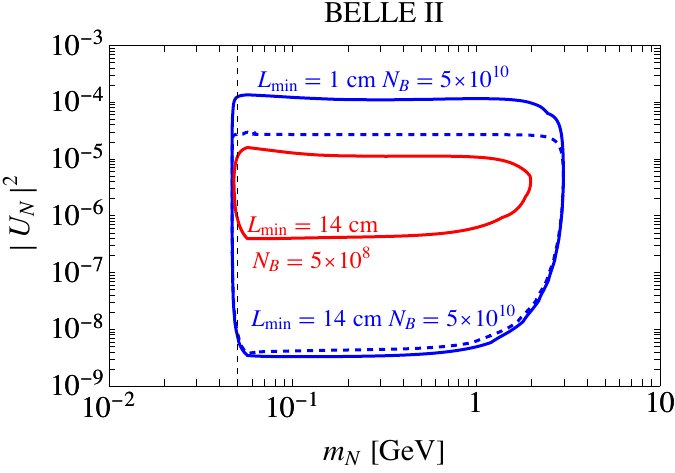}
   \caption{Projected sensitivity of Belle and Belle II in the $(m_N,|U_N|^2)$ plane, for different assumptions on the minimum resolvable displacement $L_{\rm min}$ and the number of produced $B$ mesons, $N_B$. The blue solid contour represents the most optimistic Belle II scenario, with $L_{\rm min}=1~{\rm cm}$ and $N_B=5\times10^{10}$. The blue dashed contour shows the sensitivity for the same statistics but with a more conservative displacement requirement, $L_{\rm min}=14~{\rm cm}$. The red contour corresponds to the latter displacement requirement with a smaller sample of $N_B=5\times10^8$, representative of the Belle dataset.}

    \label{fig:bbelleii}
\end{figure}

\paragraph{Belle and Belle II:} 
The Belle experiment~\cite{Belle:2024dhj} has collected $N_B=5\times10^{8}$ $B$ mesons, with an expected full luminosity reach for Belle II of $N_B=5\times10^{10}$~\cite{Belle-II:2018jsg}. We consider the decay volume as the space between the silicon vertex detector and the electromagnetic calorimeter, giving $14~{\rm cm}<L<1.55~{\rm m}$~\cite{Belle-IISVD:2022upf,Wang:2026qkv}, albeit an improvement allowing $L_{\rm min}=1$ cm may be achievable~\cite{Ferber:2022ewf,Jaeckel:2023huy}~\footnote{Tagging of displaced muons with $L\sim 1$ cm has been recently considered for muon tracks~\cite{Ferber:2022ewf,Jaeckel:2023huy}; a detailed assessment of the feasibility of the optimistic $L_{\rm min}=1~{\rm cm}$ benchmark, including realistic vertex resolution and photon reconstruction efficiencies, is left for future work.}. The geometrical acceptance of the barrel detector is $\epsilon_{\rm geo}\simeq0.74$. In \cref{fig:bbelleii} we show the reach by assuming the Belle statistics and $L_{\rm min}=14$ cm, and with the full luminosity and reduced $L_{\rm min}$, as solid and dashed red lines respectively. The latter is the one used as a red line in \cref{fig:Results} to show the projected Belle II reach.

\paragraph{LEP:} The LEP collider collected a total of $N_Z=1.7\times10^7$ $Z$ bosons~\cite{ALEPH:2005ab}. For DV searches, we consider the DELPHI detector~\cite{DELPHI:1996qcc}, with $\epsilon_{\rm geo}\simeq0.71$ and $5~{\rm cm}<L<2~{\rm m}$. The single photon triggers required $E_\g>2.5$ GeV and had an efficiency that increases with energy, from $\sim52\%$ at 6 GeV to more than $90\%$ above 10 GeV. The reach is shown as a gray region in \cref{fig:Results}; note that in the ISS model the LEP reach is severely limited to a narrow region where the branching ratio of $\phi$ to photons dominates.

\paragraph{FCC-ee:} The FCC-ee is expected to collect $N_Z=6\times10^{12}$ $Z$ bosons at full luminosity, dramatically increasing the reach of LEP. Furthermore, the IDEA detector concept~\cite{IDEAStudyGroup:2025gbt} is planned to have a coverage of $\epsilon_{\rm geo}\simeq0.99$, and a photon trigger efficiency approximately of 100\%\footnote{This efficiency is expected to drop close to the kinematic limit, $m_N\sim m_Z$, although we do not take this into consideration here.}. We consider the reference values $1~{\rm cm}<L<2~{\rm m}$ for the DV search volume, and show the projected reach as an orange line in \cref{fig:Results}.

%
\section{Conclusions}
\label{sec:Conclusions}
%
In this work we have studied Neutrino Transition Polarizability, where a light pseudoscalar $\phi$ couples active neutrinos to heavy neutral leptons and interacts with two photons. We have considered both an effective description, \cref{eq:EffectiveNTP}, and an explicit realization within the inverse seesaw mechanism. In the latter, the transition arises from the same active--sterile mixing involved in neutrino mass generation, connecting the phenomenology of the new interaction to the small breaking of the lepton number.

The main phenomenological consequence follows from the presence of $\phi$ as a light propagating degree of freedom. For $m_N>m_\phi$, the two-body decay $N\to\nu\phi$ can dominate over the weak decay channels, shortening the heavy neutrino lifetime by many orders of magnitude. The subsequent decay $\phi\to\gamma\gamma$ then produces a distinctive visible signature. In particular, even when $N$ decays promptly, the propagation of $\phi$ can lead to a macroscopically displaced photon pair. This changes the experimental strategies relevant for heavy neutrino searches: conventional downstream searches lose sensitivity, while production and decay within the instrumented detector volume become particularly promising. Below the $m_\phi$ threshold, the longer lifetime of $N$ instead favors invisible signatures and searches for kinematic modifications in meson decays.

For the reference benchmark $m_\phi=50~{\rm MeV}$ and $g_\gamma=10^{-4}~{\rm GeV}^{-1}$, we have explored visible and invisible signatures; the results are shown in \cref{fig:Results}. Displaced photons can be searched for in neutrino, flavor and collider experiments. At the DUNE Near Detector, coherent electromagnetic production provides a characteristic topology containing a primary photon and a displaced photon pair, shown in \cref{fig:ntp}. Up-scattering through weak currents supplies an additional production channel when active--sterile mixing is present, although without a primary vertex; we assume optimistically that a couple of photons with a common reconstructed vertex does not have a sizeable SM background, as $m_\phi$ is far enough from the neutral pion mass $m_{\pi^0}$. In both cases, the combination of the flux shape and the cross section scaling with $m_N$ (\cref{fig:nuxsplot}) allows DUNE to probe masses up to ${\cal O}(1)$ GeV. 

Larger masses can be explored by meson and bosons decays. Complimentary searches in $B$ meson decays can be performed at Belle II and LHCb, where the latter demonstrates larger sensitivity with respect to the former thanks to its higher statistics. LHCb can furthermore extend the mass reach by searching for displaced $W$ boson decays; in the future, the most sensitive probe will be provided by the large sample of $Z$ bosons expected at FCC-ee, making this environment especially promising.

The comparison between the effective and inverse seesaw descriptions also illustrates the importance of the underlying model. In the inverse seesaw realization, mixing-induced production can dominate, but the invisible decay $\phi\to\nu\nu$ suppresses the photon signal at large mixing. In the effective NTP scenario this channel is absent, while production in meson and electroweak boson decays requires an additional $\phi$ in the final state. Furthermore, requiring small radiative corrections to the light neutrino mass constrains the Majorana realization of the effective model; this condition is absent for a lepton number conserving Dirac state and is protected by the approximate lepton number symmetry in the inverse seesaw.

The NTP and its ISS realization indicate that simple extensions of the neutrino sector with a light dark sector can generate small neutrino masses while leading to a qualitatively new phenomenology. Our results thus aim to motivate dedicated searches for displaced vertices at present and future experiments. 


%
\section*{Acknowledgments}
We are grateful to Sam Carey, Alexander Heidelbach, Diego Redigolo and Jure Zupan for helpful discussions and insights. We also acknowledge the delightful hospitality of the Galileo Galilei Institute for Theoretical Physics, where most of this work was carried out. M.T. acknowledges support by Next Generation EU, as part of Piano Nazionale di Ripresa e Resilienza (PNRR), Missione 4, Componente 2, Investimento 1.2 - CUP I13C25000150006.

\appendix

\section{Other benchmarks}\label{sec:otherbenchmarks}
In this Appendix we briefly discuss two different benchmarks for the $\phi$ mass and photon coupling. 
\begin{figure}[t!]
    \centering
    \includegraphics[width=0.49\linewidth]{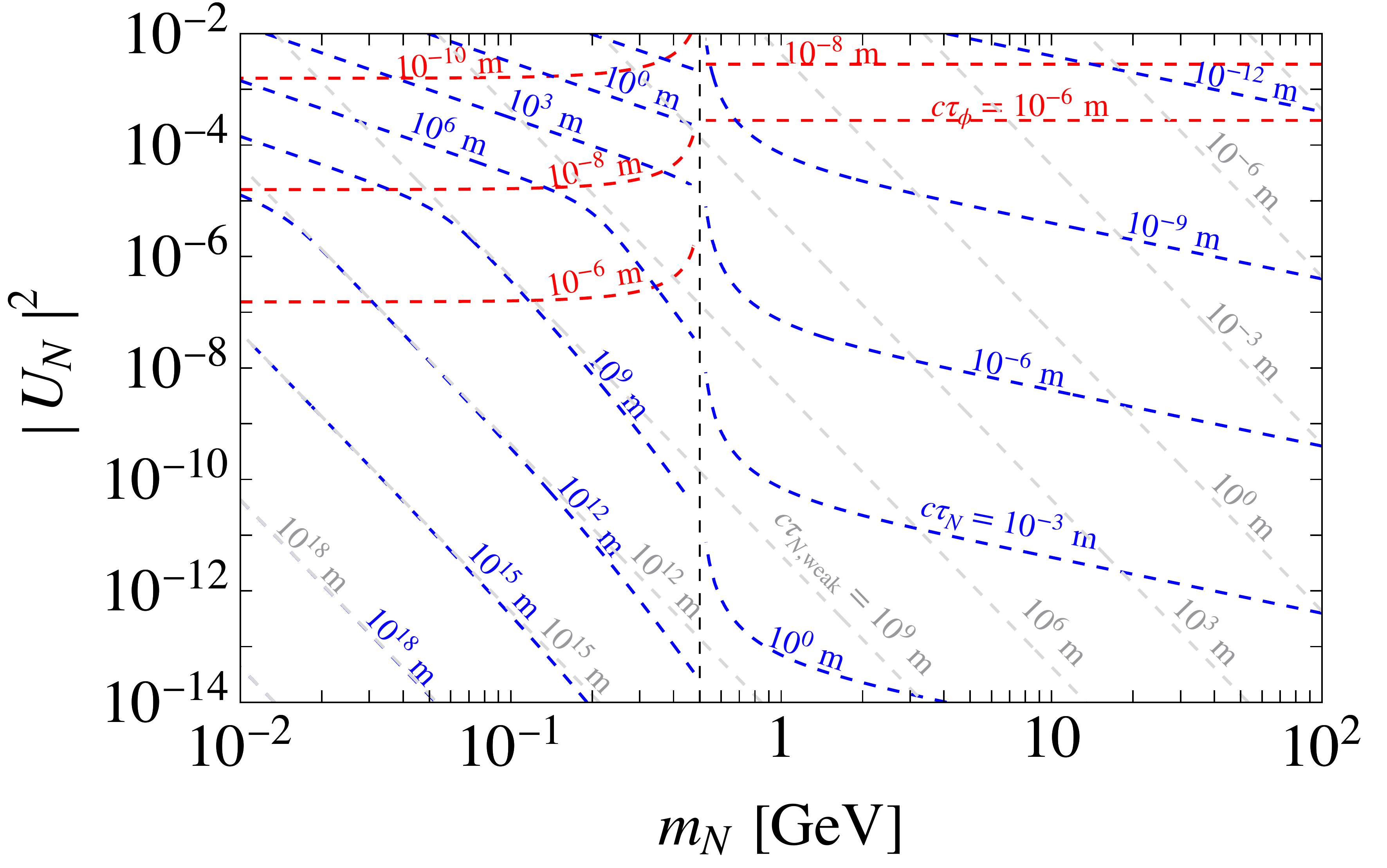}
    \includegraphics[width=0.49\linewidth]{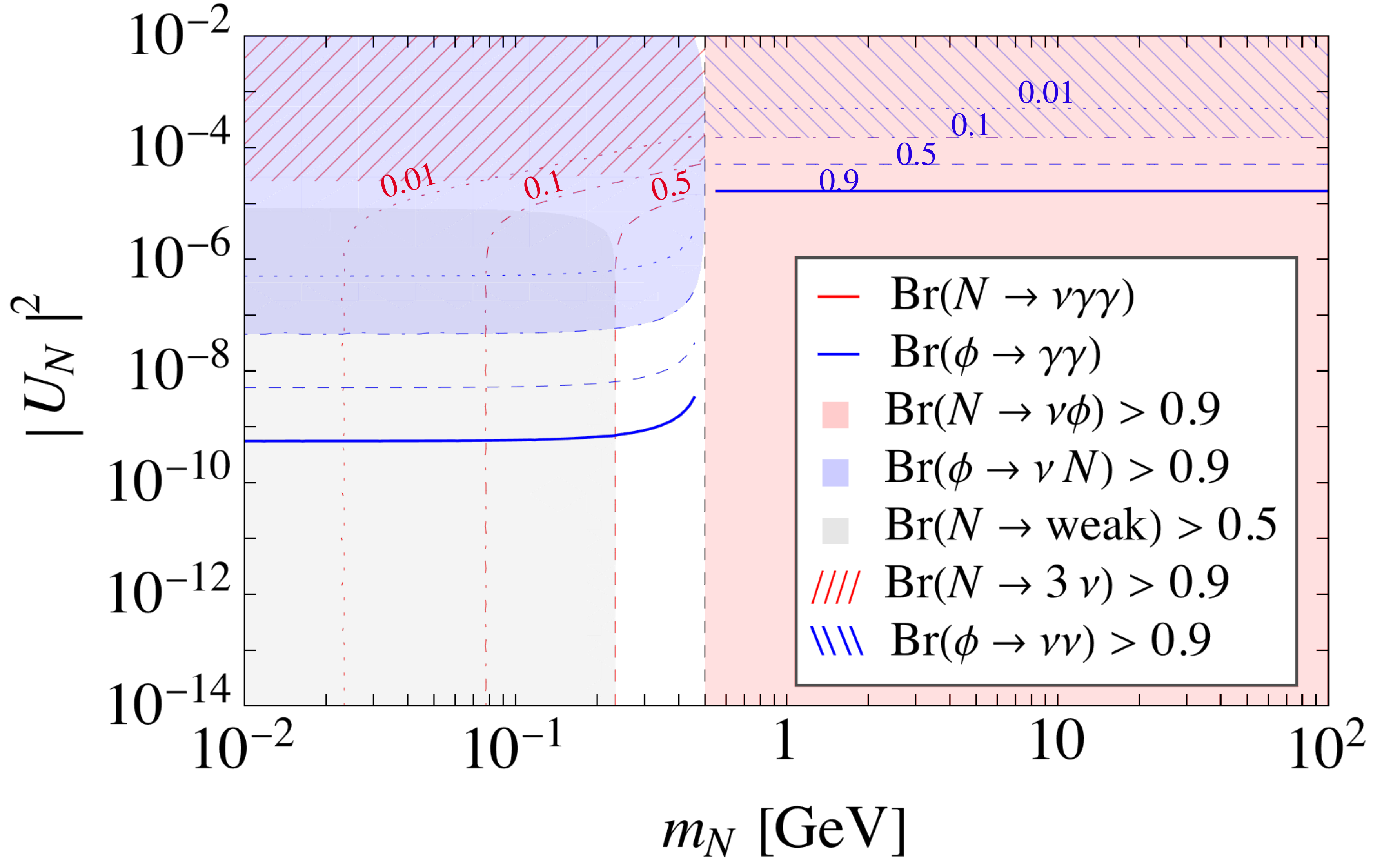}
    \includegraphics[width=0.49\linewidth]{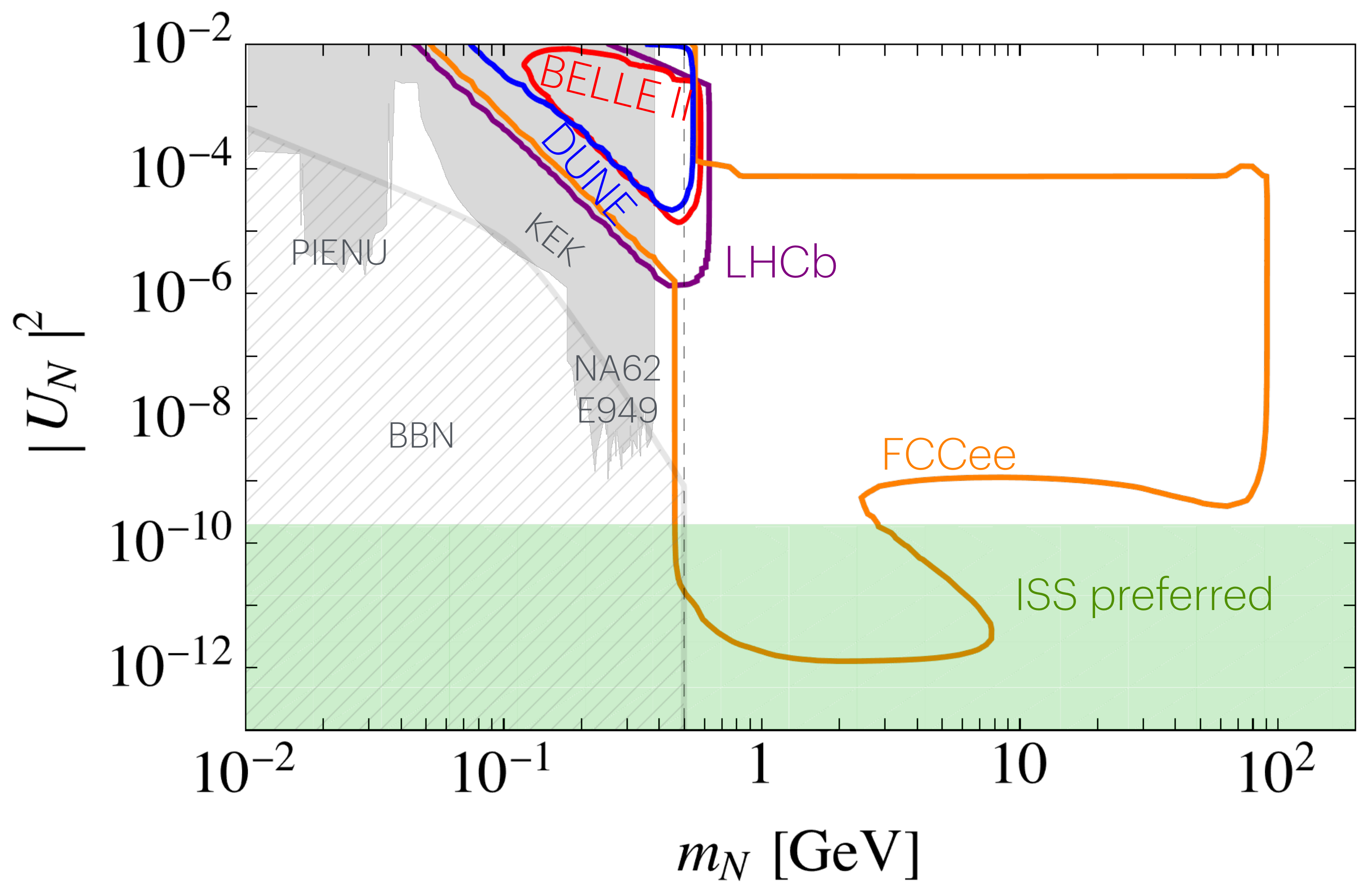}
    \includegraphics[width=0.49\linewidth]{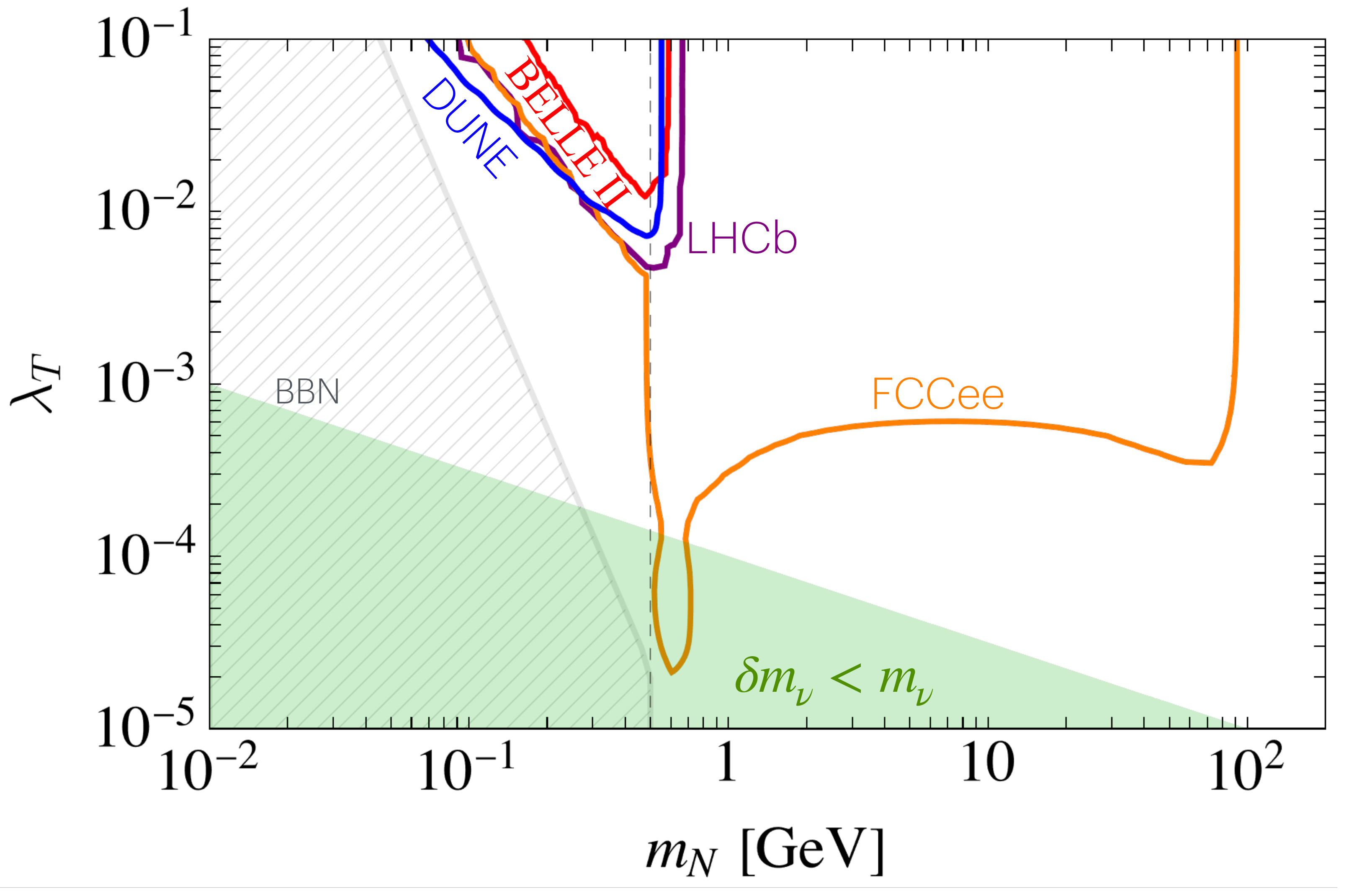}
    \caption{Same as \cref{fig:lifetimes} ({\bf top}) and as \cref{fig:Results} ({\bf bottom}), for the $m_\phi=500$ MeV benchmark.}
    \label{fig:results500}
\end{figure}

\paragraph{Heavy pseudoscalar:}
We first consider a heavier pseudoscalar, $m_\phi=500$ MeV, keeping the photon coupling fixed to the value in \cref{eq:benchmarks}. The upper panels of \cref{fig:results500} show the proper decay lengths and branching fractions of $N$ and $\phi$. Below the $m_\phi$ threshold, the larger mediator mass suppresses the radiative decay $N\to\nu\gamma\gamma$, increasing the relative importance of weak decays, as expected from \cref{eq:Case2:NTPoverMixing}. Above threshold, $N\to\nu\phi$ becomes dominant, substantially reducing the HNL lifetime, while the heavier pseudoscalar decays much more promptly, as follows from \cref{eq:phitoggdecayrate}.
In this regime, both particles typically decay promptly, except at sufficiently small mixing, where the HNL can remain long-lived, leading to less effective displaced probes. A dedicated analysis of prompt signatures, including the corresponding backgrounds, is required to assess the sensitivity in this regime and is left for future work.

The lower panels of \cref{fig:results500} show the corresponding constraints and projected sensitivities in the two scenarios, in analogy with \cref{fig:Results}. The heavier mediator generally weakens the reach of DUNE, Belle II, and LHCb, particularly above threshold, where the shorter $\phi$ decay length suppresses the rate of displaced photon signals. 
FCC-ee retains sensitivity over a much broader $m_N$ range thanks to its large $Z$ boson sample, although its reach is significantly reshaped. In particular, prompt $\phi$ decays reduce the sensitivity at larger mixing, while displaced $N$ decays can still be probed at sufficiently small mixing, where the $N$ lifetime becomes sizable.

\begin{figure}[t!]
    \centering
    \includegraphics[width=0.7\linewidth]{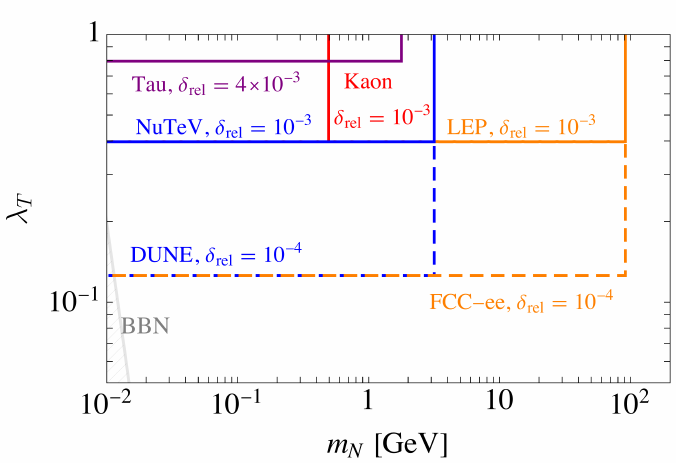}
    \caption{Schematic constraints and projected sensitivities to the NTP coupling $\lambda_T$ for $g_\gamma=0$. Solid (dashed) lines represent constraints from existing measurements (projected sensitivities), with $\delta_{\rm rel}$ denoting the relative precision of each observable. The purple (red) line corresponds to leptonic $\tau$ ($K$) decay, while the orange line indicates the invisible $Z$ width. The blue lines correspond to the NC/CC neutrino scattering ratio at NuTeV and DUNE. The vertical boundaries indicate the approximate kinematic reach of each probe. The gray dashed region is excluded by BBN.}
    \label{fig:invisibleresults}
\end{figure}

\paragraph{Invisible pseudoscalar:}
As a second benchmark, we consider a light invisible mediator $\phi$, with $m_\phi=50$ MeV and $g_\g=0$. This scenario is closely related to the neutrino transition models studied in Refs.~\cite{Hostert:2020oui,Dentler:2019dhz,Palomares-Ruiz:2005zbh,deGouvea:2019qre}.

In this scenario, the visible signatures considered in this work vanish for $m_N>m_\phi$. Below this threshold, visible final states are possible only in the ISS realization through weak currents, although with the small branching ratios shown in \cref{fig:lifetimes}. Thus the invisible pseudoscalar can be probed only by inclusive observables: modification of the NC/CC ratio (\cref{eq:ncccmodification}), shift of SM decay widths (\cref{eq:WidthShift}) and BBN. The resulting constraints and projected sensitivities are shown in \cref{fig:invisibleresults}, where solid (dashed) lines denote existing constraints (future projections). 


%
%

\bibliographystyle{JHEP}
\bibliography{bib_nu_pol_DUNE}

\end{document}